\documentclass[prb,twocolumn,superscriptaddress,amsmath,amssymb,showpacs,floatfix,preprintnumbers,keywords]{revtex4-2}
\usepackage{amsmath}
\usepackage{amssymb}
\usepackage{graphicx}
\usepackage[export]{adjustbox}
\usepackage{epstopdf}
\usepackage{color}
\usepackage{epsfig}
\usepackage{braket}
\usepackage{amsmath}
\usepackage{stackengine}
\usepackage[colorlinks=true,citecolor=blue,linkcolor=blue]{hyperref}
\usepackage{amssymb}

\DeclareMathAlphabet{\bi}{OML}{cmm}{b}{it}

\begin{document}
\title{Confinement and edge effects on atomic collapse in graphene nanoribbons} 

\author{Jing Wang}
\email[]{wangjing@hdu.edu.cn}
\affiliation{School of Electronics and Information, Hangzhou Dianzi University, Hangzhou, Zhejiang Province 310038, China}
\affiliation{Departement Fysica, Universiteit Antwerpen, Groenenborgerlaan 171, B-2020 Antwerpen, Belgium}
\affiliation{NANOlab Center of Excellence, University of Antwerp, Belgium}

\author{Robbe Van Pottelberge}
\email[]{robbe.vanpottelberge@uantwerpen.be}
\affiliation{Departement Fysica, Universiteit Antwerpen, Groenenborgerlaan 171, B-2020 Antwerpen, Belgium}
\affiliation{NANOlab Center of Excellence, University of Antwerp, Belgium}

\author{Amber Jacobs}
\author{Ben Van Duppen}
\affiliation{Departement Fysica, Universiteit Antwerpen, Groenenborgerlaan 171, B-2020 Antwerpen, Belgium}
\affiliation{NANOlab Center of Excellence, University of Antwerp, Belgium}

\author{Francois M. Peeters}
\email[]{francois.peeters@uantwerpen.be}
\affiliation{School of Physics and Astronomy and Yunnan Key Laboratory for Quantum Information, Yunnan University, Kunming 650504, China}
\affiliation{Departement Fysica, Universiteit Antwerpen, Groenenborgerlaan 171, B-2020 Antwerpen, Belgium}
\affiliation{NANOlab Center of Excellence, University of Antwerp, Belgium}

\begin{abstract}
Atomic collapse in graphene nanoribbons behaves in a fundamentally different way as compared to monolayer graphene, due to the presence of multiple energy bands and the effect of edges. For armchair nanoribbons we find that bound states gradually transform into atomic collapse states with increasing impurity charge. This is very different in zig-zag nanoribbons where multiple quasi-one-dimensional \emph{bound states} are found that originates from the zero energy zig-zag edge states. They are a consequence of the flat band and the electron distribution of these bound states exhibits two peaks. The lowest energy edge state transforms from a bound state into an atomic collapse resonance and shows a distinct relocalization from the edge to the impurity position with increasing impurity charge.
\end{abstract}



\maketitle

\section{Introduction}\label{sec:1}

Atomic collapse is a phenomenon where for sufficiently large charge of the nuclei bound states can enter the lower positron continuum and turn into quasi-bound states~\cite{ref1,ref2,ref3,ref4}. If the bound state is empty this process of entering the negative continuum is accompanied with the production of an electron-hole pair. Due to the very large nuclear charge ($Z_e$) needed in order to induce atomic collapse it was never conclusively detected in experiments~\cite{ref5,ref6}. However, the discovery of graphene enabled researchers to approach the atomic collapse problem in a different way. It was shown that due to the enhanced Coulomb interaction in graphene, atomic collapse should occur at significantly smaller charge (i.e. $Z\approx 1$) as compared to the original predicted one of relativistic atomic physics (i. e. $Z > 137$)~\cite{ref7,ref8}. Recently, atomic collapse was detected in four distinct situations: i) with charged dimers placed on top of a graphene lattice~\cite{ref9}, ii) a vacancy charged with an STM tip~\cite{ref10}, iii) collapse induced by a sharp STM tip~\cite{ref11}, and iv) using an array of subcritical charges~\cite{ref12}.

The experimental observation of atomic collapse together with its potential use for controlling charge carriers in graphene is a major motivation to study atomic collapse in more detail, e.g. by considering different arrangement of charges and different sample size. For example in Refs.~\cite{ref13,ref14,ref15,ref16,ref17} atomic collapse was studied in the presence of a dipole like field. Atomic collapse in the presence of multiple charges of equal strength was studied in Refs.~\cite{ref12,ref18,ref19}. Atomic collapse was also investigated in different systems~\cite{ref20,ref21,ref22,ref23,ref24,ref25}, e.g. in Refs.~\cite{ref24,ref25} gapped graphene was considered. In the latter case, atomic collapse was found to be more analogous to the one predicted for relativistic atoms. Instead of the sudden appearance of atomic collapse states, as in the gapless case, in gapped graphene bound states gradually turn into atomic collapse states when entering the hole continuum. Interestingly it was theoretized that atomic collapse should also occur in gapped 1D-Dirac systems~\cite{ref26}. Such a strict 1D Hamiltonian is a rather crude approximation for graphene nanoribbons demanding for a more detailed study of the problem. 

In this paper we will consider how atomic collapse manifests itself in \emph{finite width} graphene nanoribbons. It is known that nanoribbons come in different forms. There are nanoribbons with armchair or zig-zag edges and within these two types their can be either a gap or no gap depending on the number of atomic chains. On top of that the confinement in one of the spatial directions leads to the appearance of multiple energy bands. These properties are the reason that the atomic collapse of bulk graphene will be different in graphene nanoribbons as we will show in this paper. For example, we found that the Coulomb potential results in bound states at zig-zag edges. This is unexpected in view of the Klein paradox that electrons cannot be confined by electrostatic field in zero gap graphene. 

\section{Model}\label{sec:2}

Here we use the tight binding model which includes the graphene lattice structure in contrast to the continuum model used in e.g. Ref.~\cite{ref26}. For graphene we use the following tight binding Hamiltonian 
    \begin{equation}
        \hat{H} = \sum_{\langle i,j\rangle}(t_{ij}  a_i^\dagger b_j + H.c.) + \sum_i V(\overrightarrow{r_i}^A) a_i^\dagger a_i + \sum_i V(\overrightarrow{r_i}^B)  b_i^\dagger b_i
    \end{equation}
    
The first term represents the tight-binding Hamiltonian without any external fields. The hopping parameter is given by $t_{ij}$ and for graphene we take the generally accepted value -2.8 eV for nearest neighbour hopping. The operators $a_i (a_i^\dagger)$ and $b_i (b_i^\dagger)$ create (annihilate) an electron on the $i^{th}$ site of sublattice \emph{A} and \emph{B}, respectively. The last two terms include an electrostatic potential which for our case is due to the presence of a Coulomb charge, $Ze$, which we model by a Coulomb potential $V(r) = -\beta\hbar v_F/\sqrt{r^2+r_0^2}$ with $\beta = Ze^2/\kappa\hbar v_F$ the dimensionless coupling constant, $v_F$ the Fermi velocity and $\kappa$ the effective dielectric constant. We took 
 $r_0 = 0.5$ nm as a regularization parameter which is a reasonable experimental value as was shown in Ref.~\cite{ref10}. Without such a regularization a study of atomic collapse is not possible for a point-size impurity~\cite{ref27}. In order to solve the tight-binding Hamiltonian we use the open source software \emph{pybinding}~\cite{ref28}. In all the calculations we use a broadening of $0.003$ eV and the units of the Local Density of States (LDOS) are [$eV.nm^{-2}$]. A 1000 nm long nanoribbon is used to simulate the infinite long nanoribbon. For an armchair (zigzag) nanoribbon of width 4.8 nm (5 nm) the system contains $1.8*10^6$ ($1.9*10^6$) atoms.

\section{ATOMIC COLLAPSE IN GRAPHENE}\label{sec:3}

\begin{figure}[htb!]
        \centering
        \includegraphics[width=0.75\columnwidth]{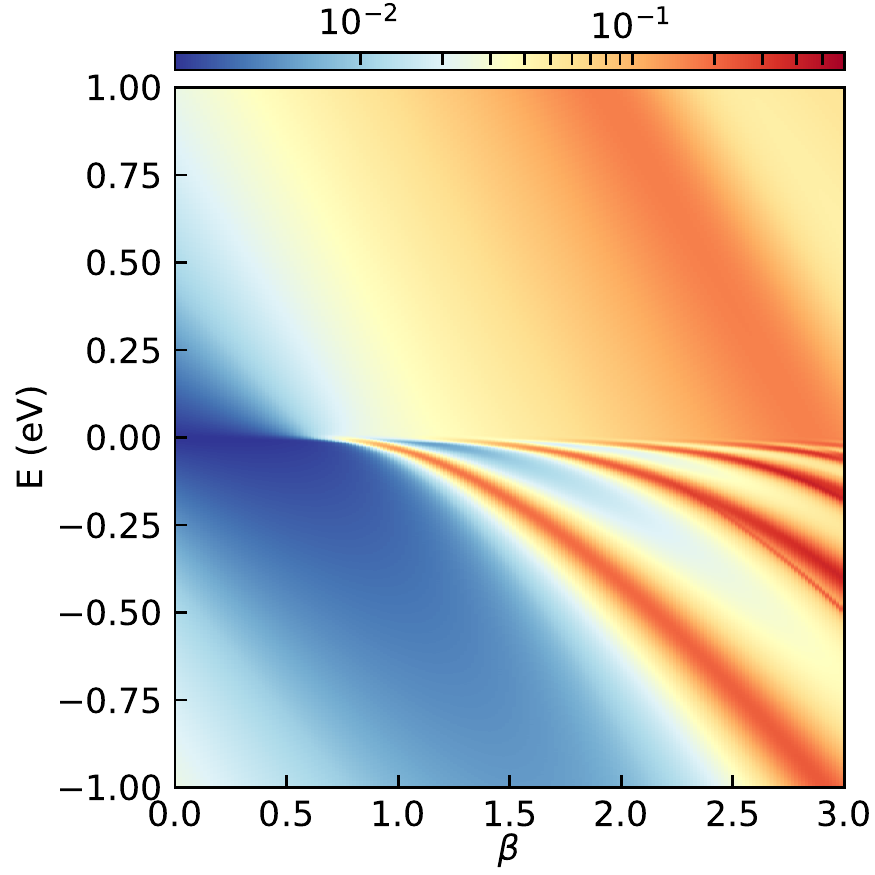}
        \caption{\label{fig:fig1} Density plot of the LDOS calculated at the impurity site as function of energy and impurity strength $\beta$ for bulk graphene.}
    \end{figure}

For completeness and for comparison purposes we review atomic collapse in graphene. In graphene the atomic collapse effect manifests itself in a different way as compared to relativistic physics. In relativistic physics bound states inside the gap region ($\Delta = 2m_0c^2$ with $m_0$ the electron rest mass and $c$ the velocity of light) decrease in energy with increasing value of the nuclear charge. However, if the nuclear charge is sufficently large the bound state(s) enters the positron continuum and hybridize with it. If this happens the bound state acquires a finite width and turns into a quasibound state which is called atomic collapse state. However, in graphene due to the gapless nature the situation is very different. This is shown in Fig.~\ref{fig:fig1} where the LDOS at the impurity site is presented as function of the charge strength $\beta$ and the energy. It can be clearly seen that when $\beta > 0.5$ resonances appear just below the Dirac point for which the LDOS exhibits peaks at the impurity site. These resonances are embedded in the hole continuum and are therefore a clear and distinct signature for atomic collapse since they represent an electron state hybridized with the negative continuum. Note that with increasing value of the charge the resonances move to lower energies and their width increases. 

In the case of graphene nanoribbons the spectrum is different in two fundamental ways: i) the single conic bands of graphene are replaced by an infinite number of bands, and ii) depending on the width of the ribbon a gap can appear in the spectrum. In the next section we will discuss how these changes in the spectrum affect the manifestion of the atomic collapse effect in graphene nanoribbons.

    \begin{figure}[htb!]
        \centering
        \includegraphics[width=0.75\columnwidth]{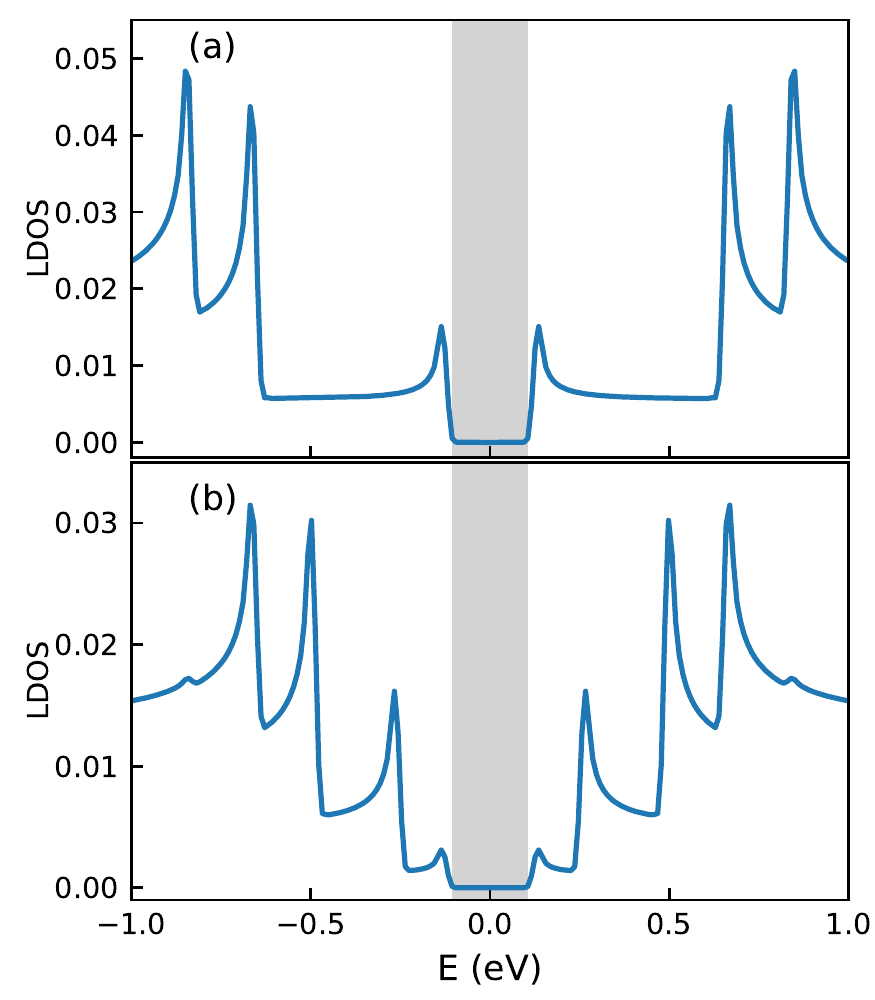}
        \caption{\label{fig:fig2} (a) LDOS measured at the center of an armchair nanoribbon of width $4.8$ nm. The gray region indicates the gap region where the LDOS is zero and consequently no states are found. (b) The same but the LDOS is calculated a distance 0.18 nm from the center of the nanoribbon. $\beta=0$ in both cases}.
    \end{figure}
    
    \begin{figure}[htb!]
        \centering
        \includegraphics[width=0.75\columnwidth]{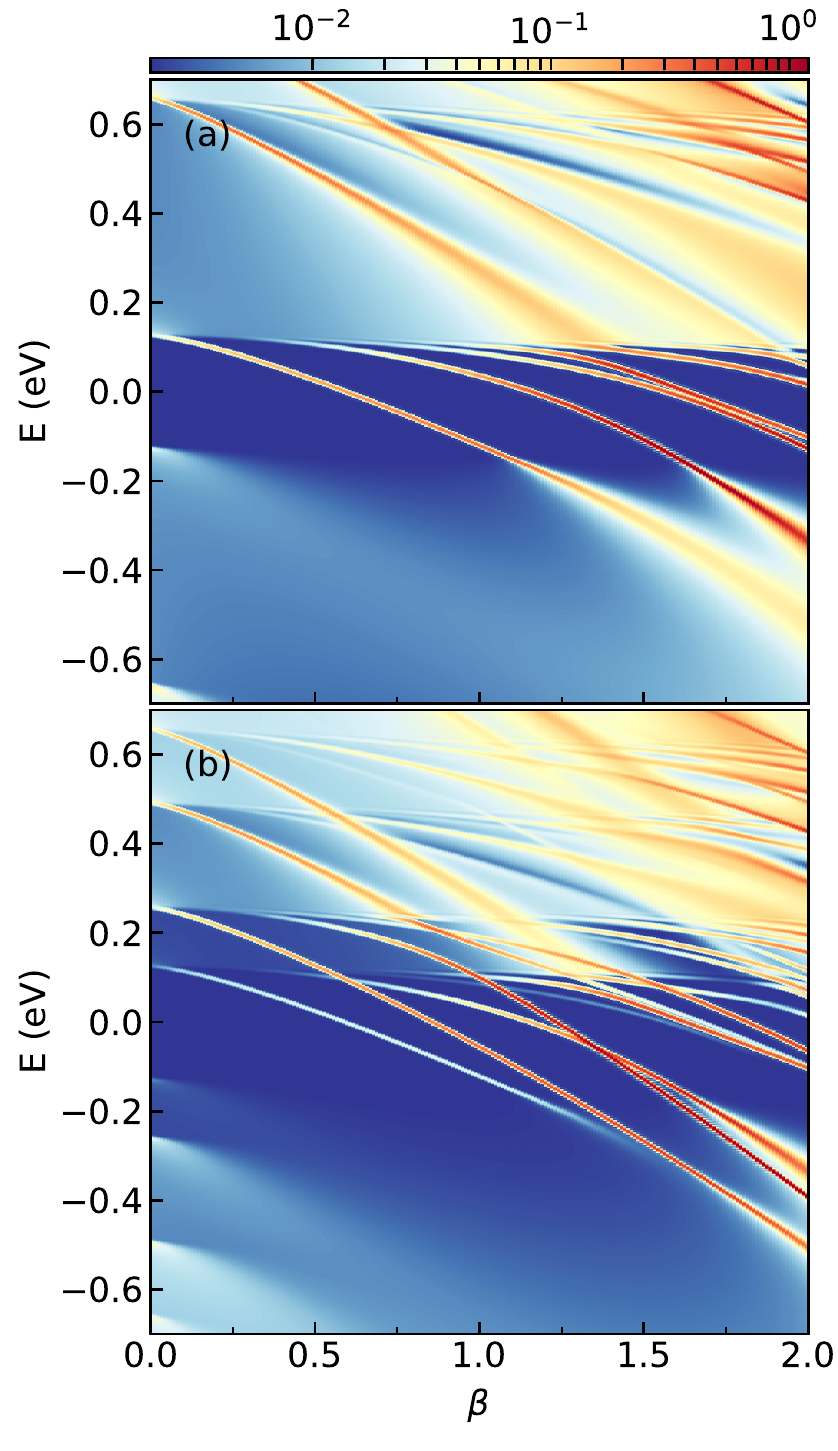}
        \caption{\label{fig:fig3} LDOS measured at the impurity position as function of the charge for a graphene nanoribbon of width $4.8$ nm, this corresponds to a nanoribbon with $N = 39$ atomic chains. In (a) the LDOS is shown for a charge placed in the center of the nanoribbon and the LDOS is measured at the impurity site while in (b) the LDOS is measured a distance 0.18 nm away from the impurity in the direction perpendicular to the nanoribbon.}
    \end{figure}

\section{ATOMIC COLLAPSE IN ARMCHAIR NANORIBBONS}\label{sec:4}
    
For armchair nanoribbons their are two major types of ribbons depending on the width of the ribbon~\cite{ref29}: i) a gap in the spectrum is found when the number of atomic chains is $N = 3p$ or $N = 3p + 1$ with $p$ a positive integer. This gap is proportional to the inverse of the ribbon width. These nanoribbons are semiconducting; ii) when $N = 3p + 2$ the spectrum of the nanoribbon is gapless and these nanoribbons are metallic. However, it was shown in experiments that all the armchair nanoribbons are semiconducting~\cite{ref30}. This discrepancy can be explained by including third nearest neighbour hopping. Since all armchair nanoribbons are gapped we will focus on nanoribbons that are gapped within a nearest neigbour hopping tight binding model. Including third nearest neigbour hopping only leads to quantitative corrections making the metallic nanoribbons very similar to the semiconducting ones. In Fig.~\ref{fig:fig2}(a) an example of a typical LDOS of an armchair nanoribbon is shown. We calculated the LDOS at the center of the armchair nanoribbon with width 4.8 nm. The gap ($\Delta = 0.25$ eV) is shown in gray. The cusps in the LDOS are typical for nanoribbons and a consequence of the 1D nature of the spectrum and correspond to the onset of a new subband. In Fig.~\ref{fig:fig2}(b) the LDOS is calculated a distance $0.18 nm$ from the center of the same nanoribbon in the direction perpendicular to the nanoribbon length. Figs.~\ref{fig:fig2}(a) and~\ref{fig:fig2}(b) illustrate that the electron probability corresponding to the different bands can be zero at some of the carbon rows.

The gap region will allow for impurity bound states. This is in contrast with gapless pristine graphene where only quasi-bound states are possible. The effect of such gap is clearly shown in Fig.~\ref{fig:fig3}(a) for -0.125 eV $< E <$ 0.125 eV where the LDOS is measured at the impurity site as function of the impurity strength $\beta$ for an impurity placed at the center of the nanoribbon. The width of the ribbon taken along $x-$direction is chosen to be 4.8 nm, using the value of the inter carbon distance $a_{cc} = 0.142$ nm this gives $N = 39$ atomic chains. When the strength of the charge is gradually increased a clear and distinct state sinks into the gap region corresponding to a bound state. The LDOS of this state increases when the charge increases which is due to the fact that the state gets localized closer to the impurity. The lowest bound state inside the gap keeps its bound state character until the charge reaches $\beta \approx 1.25$. After this the bound state gets redistributed over the negative continuum states and aquires a finite width and turns into a resonant state. Note that for larger $\beta$ more bound states appear in the gap region. All these bound states turn into resonances when they enter the negative continuum region.

In Fig.~\ref{fig:fig3}(b) we show the LDOS at a distance $0.18$ nm away from the center of the nanoribbon. Interestingly, more states and bands appear as compared to measuring the LDOS at the center of the nanoribbon. This behavior can be explained from the fact that in armchair nanoribbons some states show zero LDOS for certain rows of atoms as discussed in Ref.~\cite{ref31}. The symmetric position of the charge in the middle of the ribbon implies that some states maintain zero LDOS at the center of the nanoribbon, and consequently do not show up when calculating the LDOS at the center. This behavior is very similar to the effect of defects studied in Ref.~\cite{ref31}. In Fig.~\ref{fig:fig3}(b) we notice the appearance of an extra band (around $E \approx 0.25$ eV) with a diving series of states which qualitatively behave very similar to the states inside the gap discussed earlier: they show a similar dependence on $\beta$ and turn into quasi-bound states when entering the lower continuum. 

    \begin{figure}[htb!]
        \centering
        \includegraphics[width=\columnwidth]{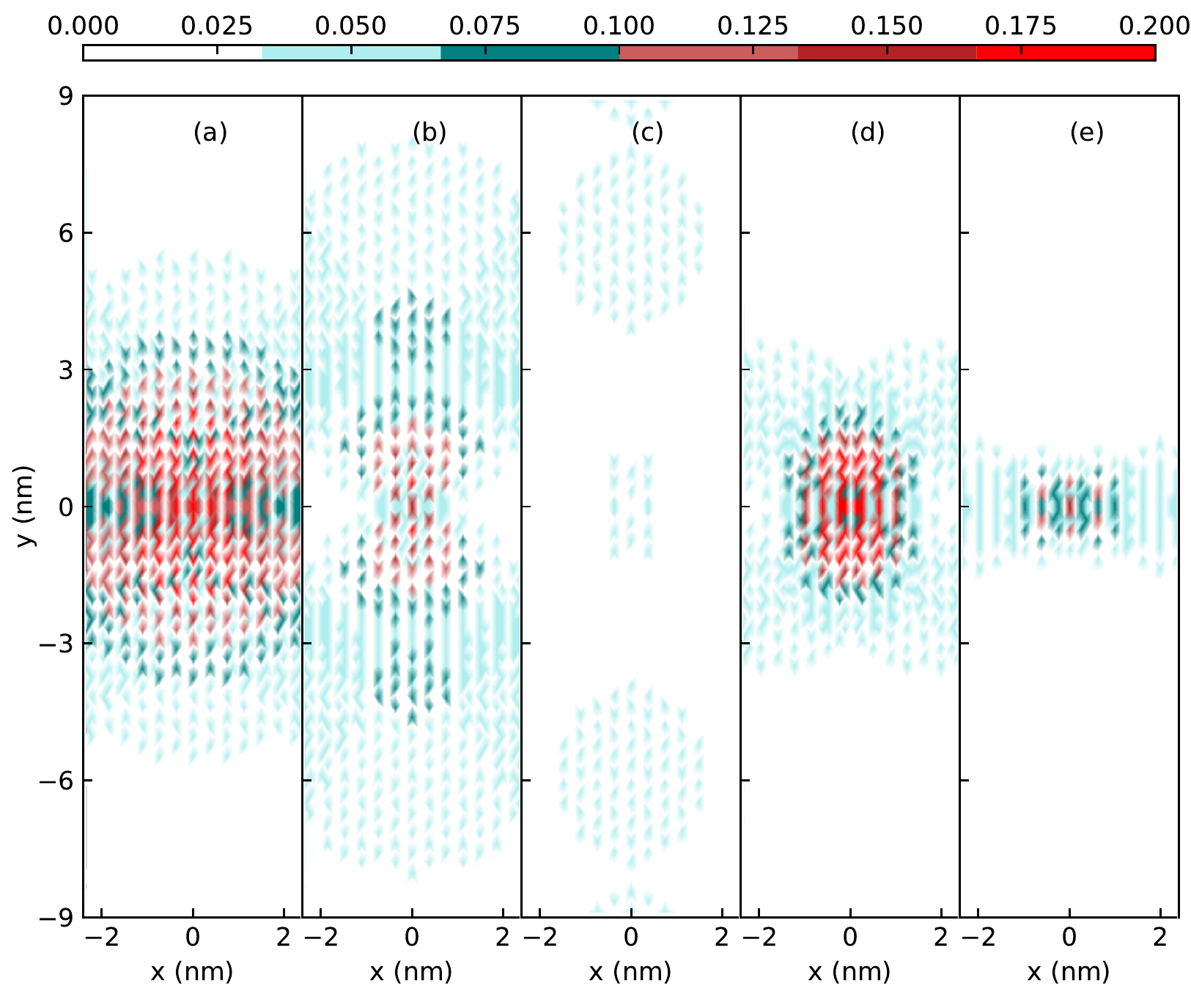}
        \caption{\label{fig:fig4} Spatial LDOS for $\beta = 1$ of the first five states (see Fig. 3(a)) with energy: (a) $E = -0.12$ eV, (b) $E = 0.037$ eV, (c) $E = 0.086$ eV, (d) $E = 0.105$ eV, and (e) $E = 0.272$ eV. Red(blue) represents high(zero) LDOS while blue represents zero LDOS.}
    \end{figure}
    \begin{figure}[htb!]
        \centering
        \includegraphics[width=\columnwidth]{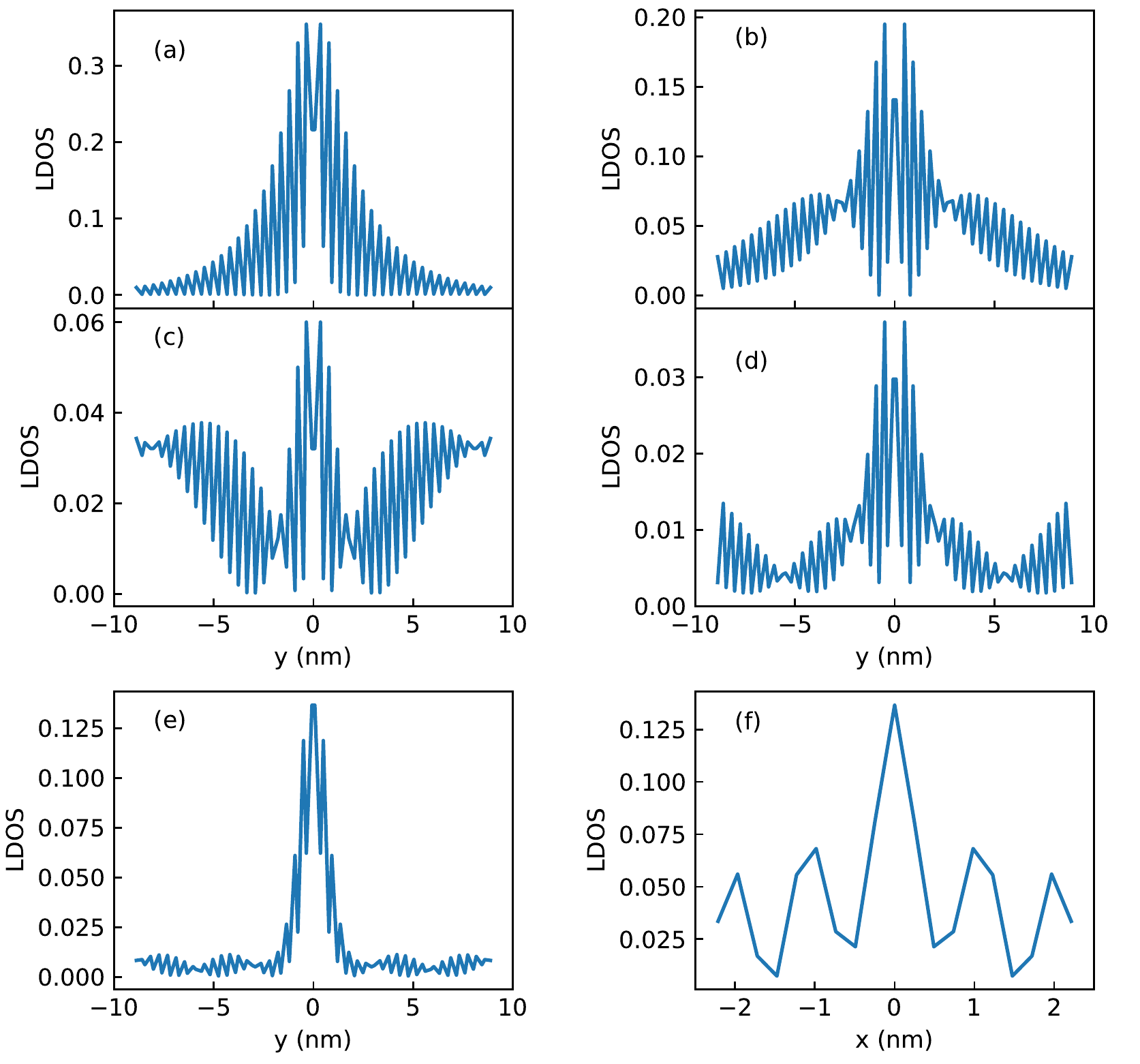}
        \caption{\label{fig:fig5} Cut through $x = 0$ in the y direction for the spatial LDOS calculations shown in Fig.~\ref{fig:fig4}. In (f) also a cut along $x$-direction for $y = 0$ is shown for the state corresponding to (e).}
    \end{figure}
So far we only studied the energy dependence of the LDOS at a single atomic position. In Fig.~\ref{fig:fig4} we plot the spatial distribution of the LDOS for the first 5 electronic states observed in Fig.~\ref{fig:fig3}(a) at $\beta = 1$. Figures 4(a)-4(d) correspond to states inside the gap while Fig.~\ref{fig:fig4}(e) corresponds to the first state outside the gap and belongs to the next subband. In contrast to Schr$\ddot{o}$dinger physics with symmetric potentials, here we do not have even and odd solutions and therefore no clear nodes in the wavefuntions (or LDOS) are found. The LDOS exhibits rather a dumbell character which is made more clear in Figs.~\ref{fig:fig5}(a) -~\ref{fig:fig5}(d) where we show cuts through the LDOS of Figs.~\ref{fig:fig4}(a) -~\ref{fig:fig4}(d). The discrete nature of the LDOS reflects the discrete graphene lattice. These states have some similarity to the ones found for a strict 1D-Dirac Hamiltonian~\cite{ref26}. 

    \begin{figure}[htb!]
        \centering
        \includegraphics[width=0.75\columnwidth]{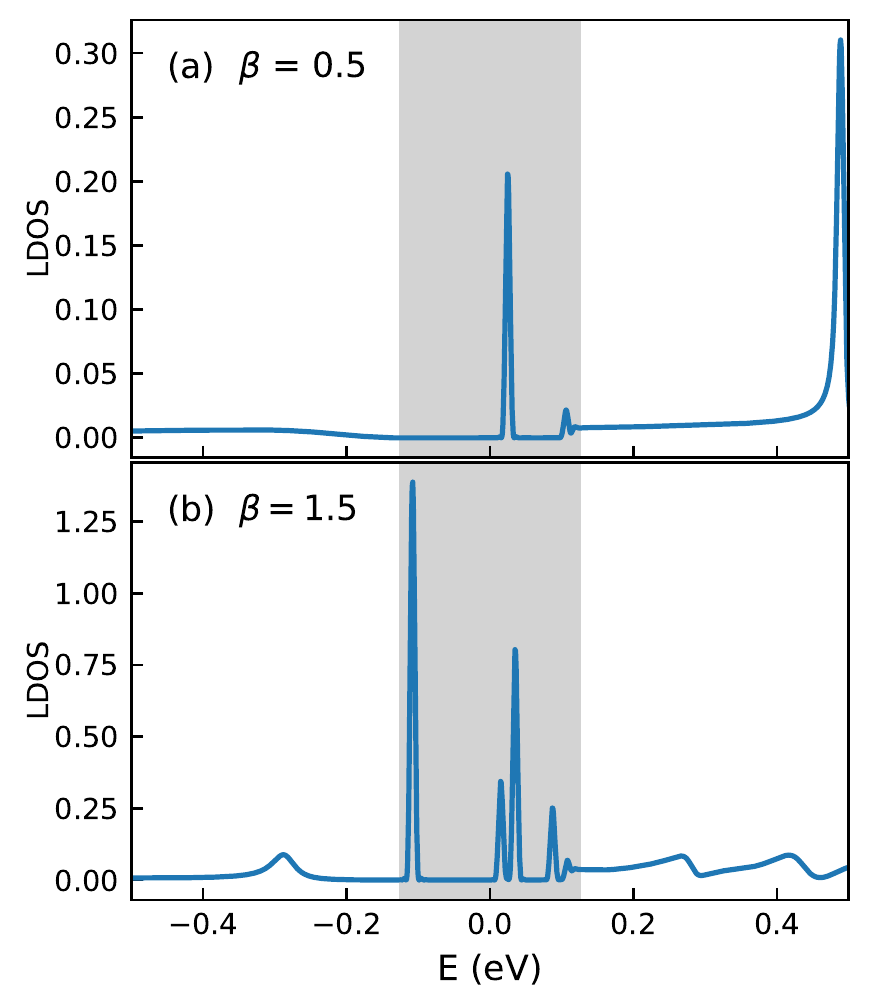}
        \caption{\label{fig:fig6} Cut of the LDOS presented in Fig. 3(a). In (a) the LDOS is shown for $\beta = 0.5$ and in (b) for $\beta = 1.5$. As in Fig. 2 the gap is indicated by the gray region.}
    \end{figure}
In Fig.~\ref{fig:fig3}, next to the states located inside the gap region quasi-bound states appear that are attached to higher energy bands. Those states can be seen in the LDOS as peaks already at small charge right below the band edge. They hybridize almost immediately with the underlying continuum acquiring a finite width. Such resonances for positive energy are not atomic collapse states. An atomic collapse state is a conduction band state redistributed over valence band states that are located in the negative continuum. The resonances observed for positive energy are therefore resonant states but are not related to atomic collapse. These hybridized states were investigated in Ref.~\cite{ref32} for small $\beta$ within a continuum model. Our tight-binding results show that these states should appear as a clear signature in LDOS measurments with e.g. a STM tip. Note that this is related to some states that were recenly predicted for bilayer graphene with a Coulomb impurity~\cite{ref33}. In Figs.~\ref{fig:fig5}(e) and~\ref{fig:fig5}(f) the spatial LDOS for the first quasi-bound state observed in the positive continuum in Fig.~\ref{fig:fig3}(a) is shown. This state shows no dumbell-like distribution but a more 1S like atomic orbital shape which is confirmed by a cut of the spatial LDOS shown in Fig.~\ref{fig:fig5}(e) for $x = 0$. Also a cut for $y = 0$ is shown next to the latter which indicates that the state is not confined along the nanoribbon. Note that in Fig.~\ref{fig:fig3}(b) the states below the second energy band do not hybridize immediately with the underlying continuum. This behavior is a consequence of the symmetric placement of the charge. Further in the manuscript the effect of an assymetricly placed charge on these states will be investigated.

    \begin{figure}[htb!]
        \centering
        \includegraphics[width=0.75\columnwidth]{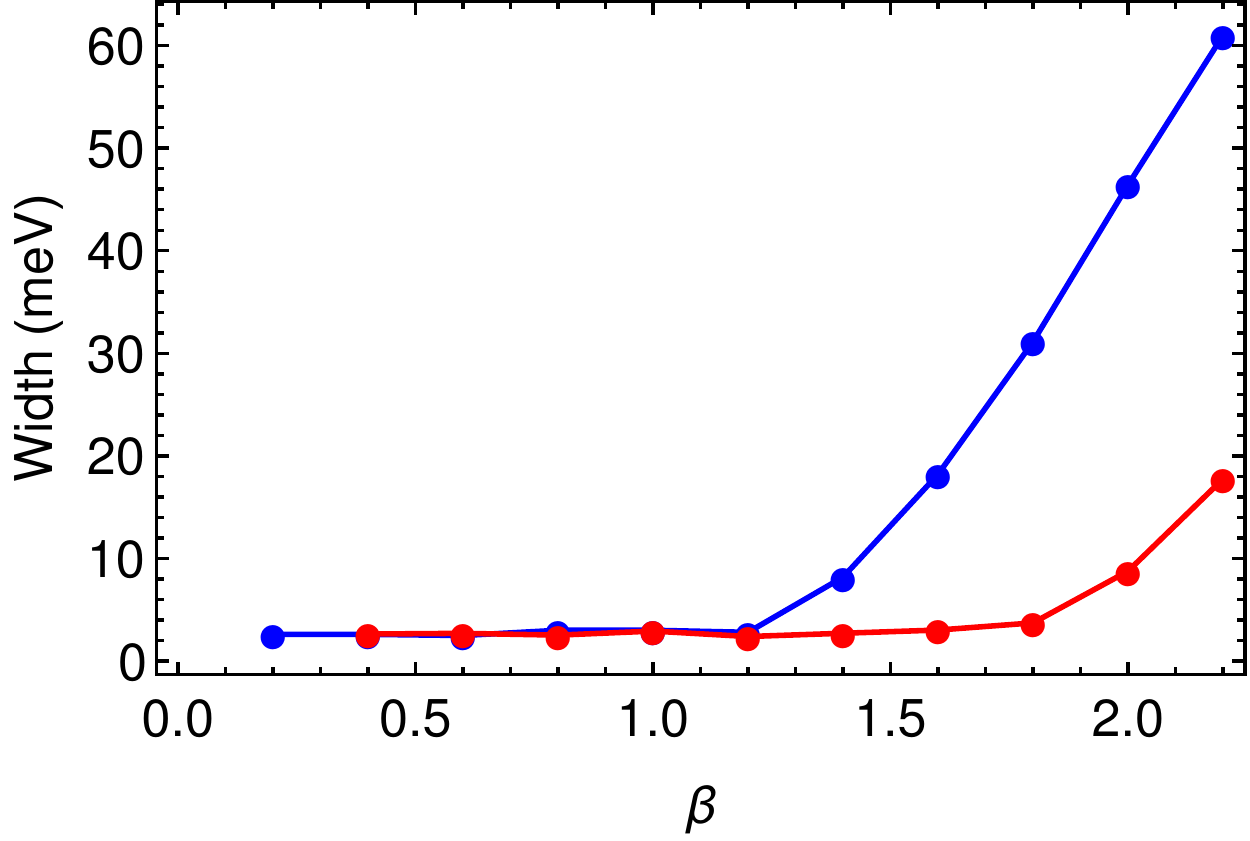}
        \caption{\label{fig:fig7} Width of first (blue) and second (red) bound state turning into a resonant state shown in Fig. 3(a) as function of the impurity strength $\beta$. The width is defined as the energy range over which the peak loses 30$\%$ of it’s intensity.}
    \end{figure}
    
In Fig.~\ref{fig:fig6} a cut of the LDOS of Fig.~\ref{fig:fig3}(a) is shown for two values of $\beta$: (a) $\beta = 0.5$ and (b) $\beta = 1.5$. For $\beta = 0.5$ the gap region can be clearly seen (denoted by the gray region). Two bound states are visible inside the gap. At the edge of the gap region a small peak can be observed which corresponds to the second bound state. When the charge is further increased to $\beta = 1.5$ the first bound state of $\beta = 0.5$ has now entered the negative continuum turning into a quasi-bound state with a sizeable width. Inside the gap region an additional number of bound states appear due to the increased value of the charge. For energy $E > 0.125$ eV two resonant states originating from the first energy band can be observed.

In order to clearly show the bound state to quasi-bound state transition seen in Fig.~\ref{fig:fig3}(a) we plotted the broadening of the first and second bound state as function of the strength of the charge in Fig.~\ref{fig:fig7}. For $0 < \beta < 1.2$ the width of the first bound state remains clearly constant and small ($\approx 0.003$ eV which is the broadening used in the calculation of the LDOS) which is representative for a bound state. However after $\beta \approx 1.2$ the width starts to increase drastically signifying the transition from a bound state to a quasi-bound state. The second bound state (red curve in Fig.~\ref{fig:fig7}) turns into an atomic collapse state at $\beta \approx 1.8$. After the bound state has entered the continuum its width increases with decreasing energy similar as in the case of bulk graphene shown in Fig.~\ref{fig:fig1}. The results in Fig.~\ref{fig:fig7} thus show a distinct bound state to atomic collapse state transition.

    \begin{figure}[htb!]
        \centering
        \includegraphics[width=0.75\columnwidth]{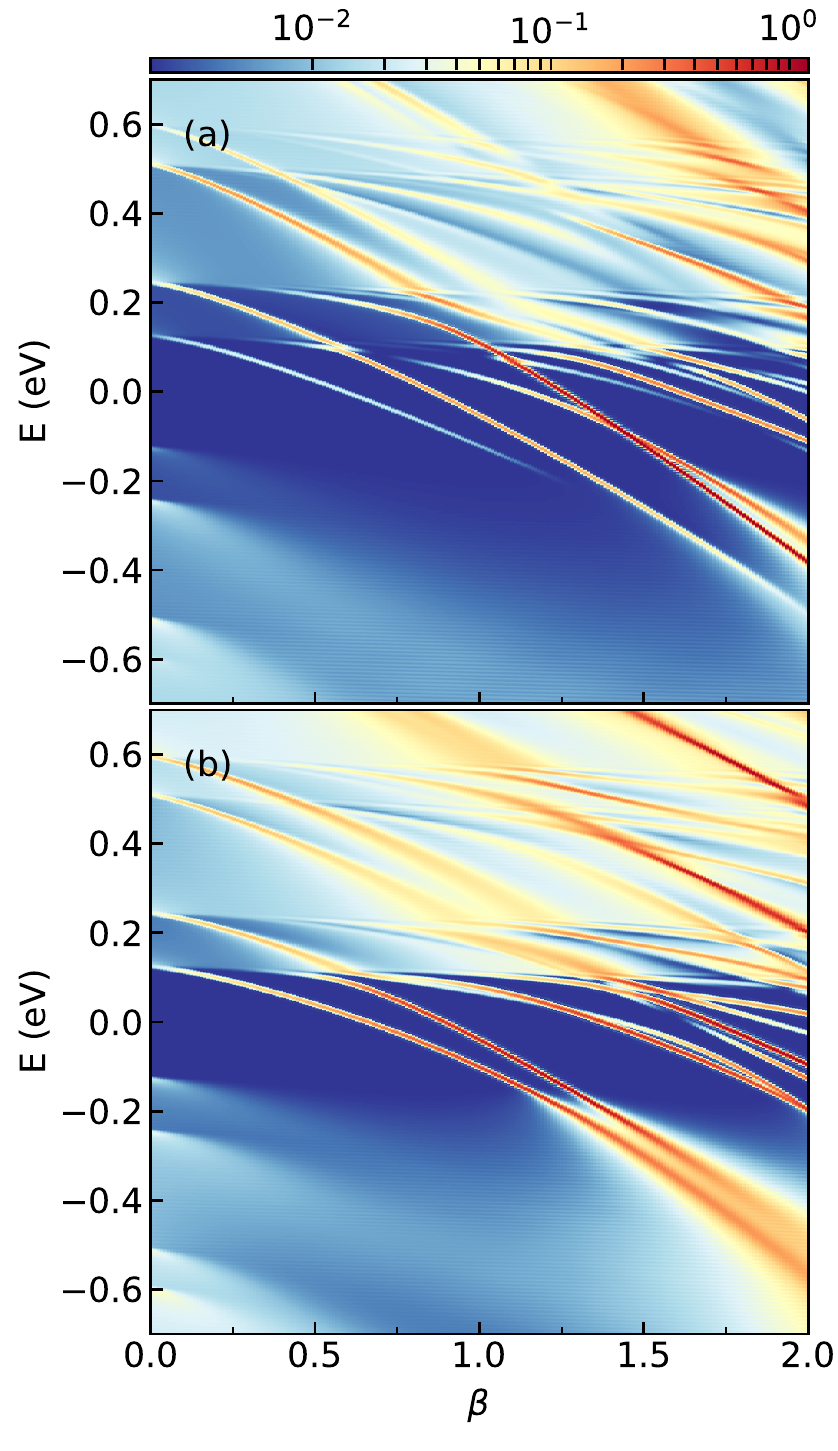}
        \caption{\label{fig:fig8} LDOS measured at the impurity position as function of the charge for a graphene nanoribbon of width 4.8 nm. In (a) the charge is placed a distance 0.5 nm from the center of the ribbon while in (b) the charge is placed a distance 2 nm away from the center.}
    \end{figure}

In all the above discussion, the charge was considered to put at the center of the nanoribbon. The LDOS measured at the impurity show less peaks than it was measure 0.18 nm away from the impurity in Fig.~\ref{fig:fig3}. This phenomenon triggers us to study the effect of the position of the charge on the spectrum. The corresponding LDOS is shown in Fig.~\ref{fig:fig8} for the same nanoribbon as the one in Fig.~\ref{fig:fig3} but now for a charge placed 0.5 nm from the center of the nanoribbon in Fig.~\ref{fig:fig8}(a) and 2 nm from the center of the nanoribbon in Fig.~\ref{fig:fig8}(b). For a small asymmetric placement of the charge the spectrum looks very similar to the one shown in Fig.~\ref{fig:fig3}(b) for a symmetric placement but where the LDOS is measured away from the charge position. However, the bound states originating from the second energy band at $E \approx 0.22$ eV in Fig.~\ref{fig:fig8}(b) start to show hybdridization with the underlying continuum acquiring a finite width. This confirms our theory that the special non-hybridizing behavior of these states discussed in connection with Fig.~\ref{fig:fig3}(b) is due to the symmetric placement of the charge. Increasing the asymmetry even further as shown in Fig.~\ref{fig:fig8}(b) we notice that these states show an even stronger hybridization making our point even stronger. Qualitatively similar features in the spectrum are seen when placing the charge asymmetrically. All the general features discussed for the symmetric case remain: i) states inside the gap show a transition from bound to atomic collapse state with increasing charge $\beta$, ii) multiple energy bands appear, and iii) below these higher energy bands states appear that almost immediately hybridize with the underlying positive continuum. This shows that the physics presented in this manuscript should be robust in experiments almost independent of the exact position of the charge, paving the way for the first experimental observation of a bound state to atomic collapse state transition.

\section{ZIG-ZAG NANORIBBONS}\label{sec:5}
    \begin{figure}[htb!]
        \centering
        \includegraphics[width=0.75\columnwidth]{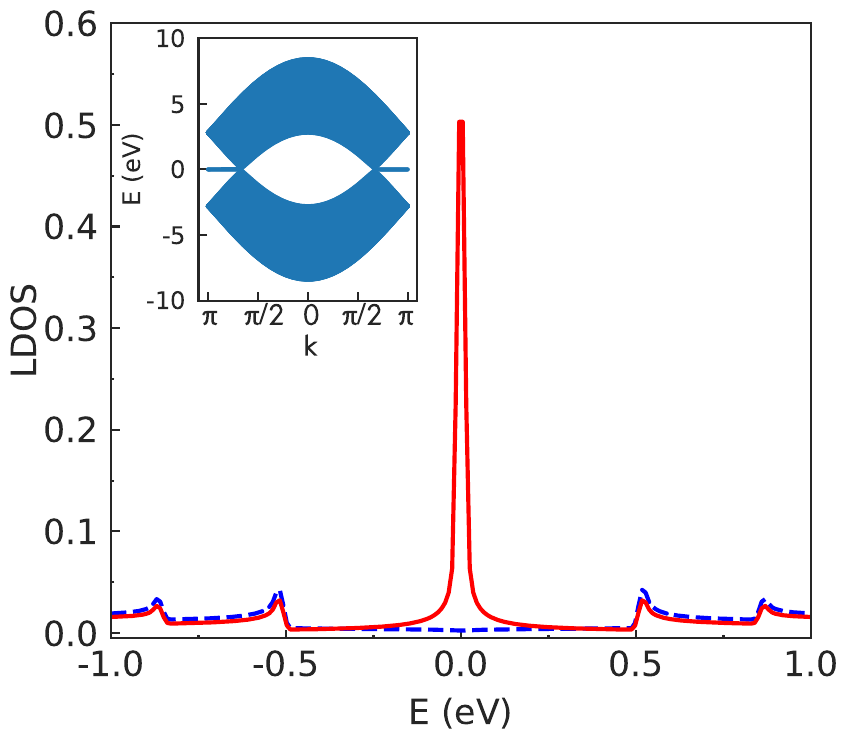}
        \caption{\label{fig:fig9} LDOS measured at the center (dashed blue) and at 2 nm from the center (solid red) of a zig-zag nanoribbon of width 5 nm without an impurity. The insert is the band structure of the zig-zag ribbon.}
    \end{figure}
    \begin{figure}[htb!]
        \centering
        \includegraphics[width=0.75\columnwidth]{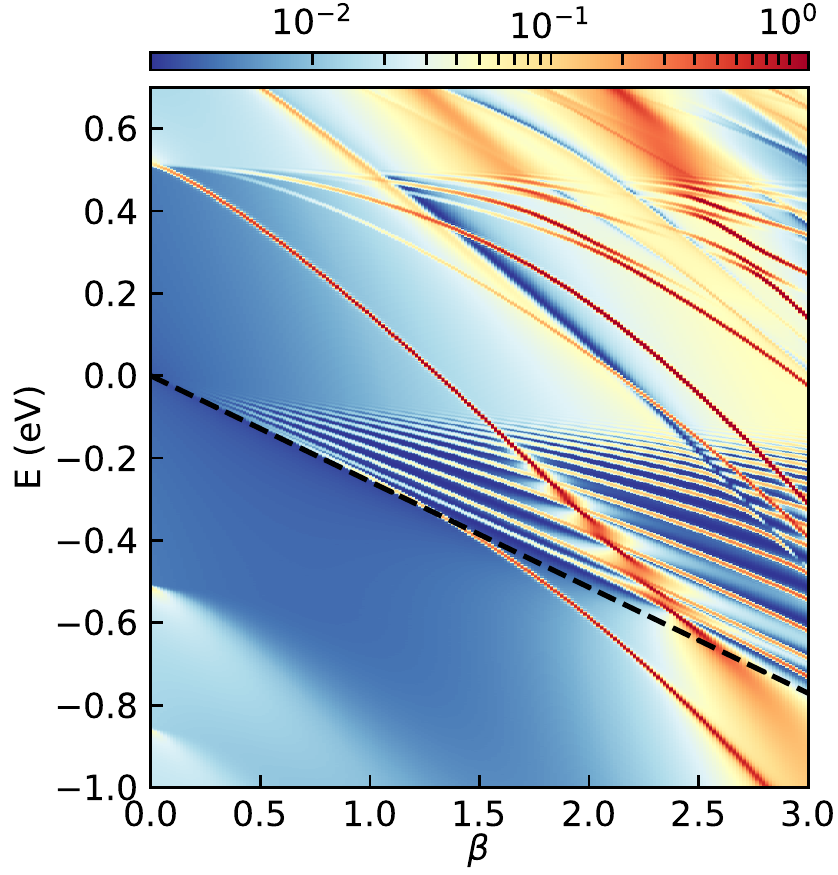}
        \caption{\label{fig:fig10} LDOS measured at the impurity site as function of the charge strength $\beta$ and energy for a graphene nanoribbon with zig-zag edges of width 5 nm. The impurity is located at the center of the nanoribbon. The dashed line is the value of the Coulomb potential at the edge.}
    \end{figure}
 
In the case of zig-zag nanoribbons there is an extra element that is different from armchair nanoribbons, namely the presence of zero energy edge states which have a flat band character. It is expected that these edge states will be strongly influenced by the presence of the impurity. In Fig.~\ref{fig:fig9} we show the LDOS calculated at the center of a zigzag nanoribbon (blue) of width 5 nm without an impurity. The cusps signifying the multiple energy bands can be clearly seen. Note that compared to the LDOS of the armchair nanoribbon discussed in the previous section the LDOS does not show a gap around $E = 0$ eV. When the LDOS is calculated 2 nm away from the center of the nanoribbon (red) an extra peak at $E = 0$ emerges due to the edge states.

    \begin{figure}[htb!]
        \centering
        \includegraphics[width=0.75\columnwidth]{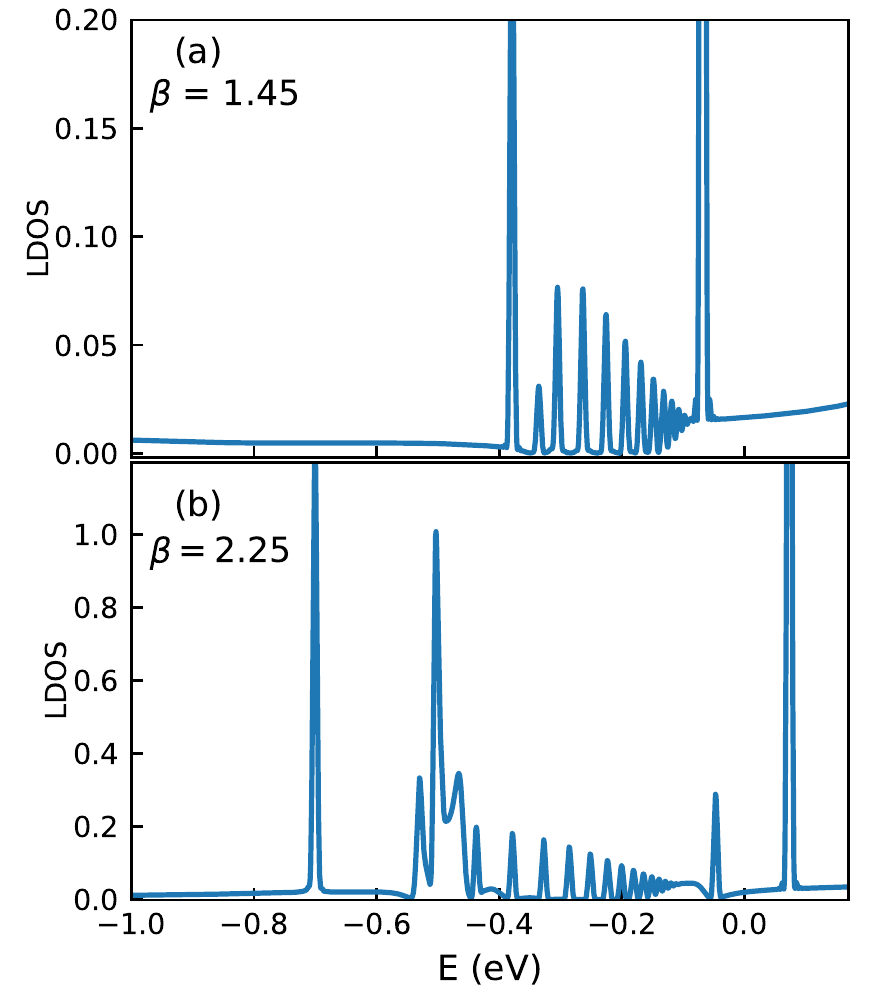}
        \caption{\label{fig:fig11} Cut of the spatial LDOS shown in Fig. 10 for (a) $\beta$ = 1.45 and (b) $\beta$ = 2.25.}
    \end{figure}

In Fig.~\ref{fig:fig10} the energy dependence of the LDOS is shown as function of the impurity strength for a zig-zag nanoribbon of width 5 nm for an impurity placed at the center of the nanoribbon. A number of states originating from $E = 0$ eV at $\beta = 0$ can be clearly observed. These states are pulled towards lower energy with increasing impurity charge. Since they originate from $E = 0$ eV this band of discrete states can be attributed to the edge states. They are bound states which we confirmed by the fact that their width increases linearly with the imposed numerical broadening and their position and width did not change when we increase the length of the graphene nanoribbon. The lowest state shows interesting behavior with increasing impurity charge. It starts as a bound state which can be seen from the narrow width of the LDOS. However, when the impurity charge increases the state comes in contact with the lower continuum band (starting at $E \approx$ -0.50 eV) and gradually turns into an atomic collapse resonance. The width of this resonance increases with increasing charge. The reason that the band of edge states can support bound states lies in the fact that this band consists of a mixture of conduction and valence states. Consequently the conduction band nature leads to the appearance of the previously discussed bound states. The behavior of the states in the positive energy range of Fig.~\ref{fig:fig10} is similar to the ones shown for an armchair nanoribbon in Fig.~\ref{fig:fig3} and therefore will not need any further discussion. Just below $E \approx 0.5$ eV the bound states do not hybridize with the continuum below. However, when the charge of the impurity increases these bound states come into contact with the band of edge states gradually turning into a resonance which is modulated by the appearance of the edge states. This behavior can be most profoundly seen for the first state originating from $E \approx 0.5$ eV. After being modulated by the edge state the quasi-bound nature of the state becomes clear around $\beta \approx 2.5$.

    \begin{figure}[htb!]
        \centering
        \includegraphics[width=0.75\columnwidth]{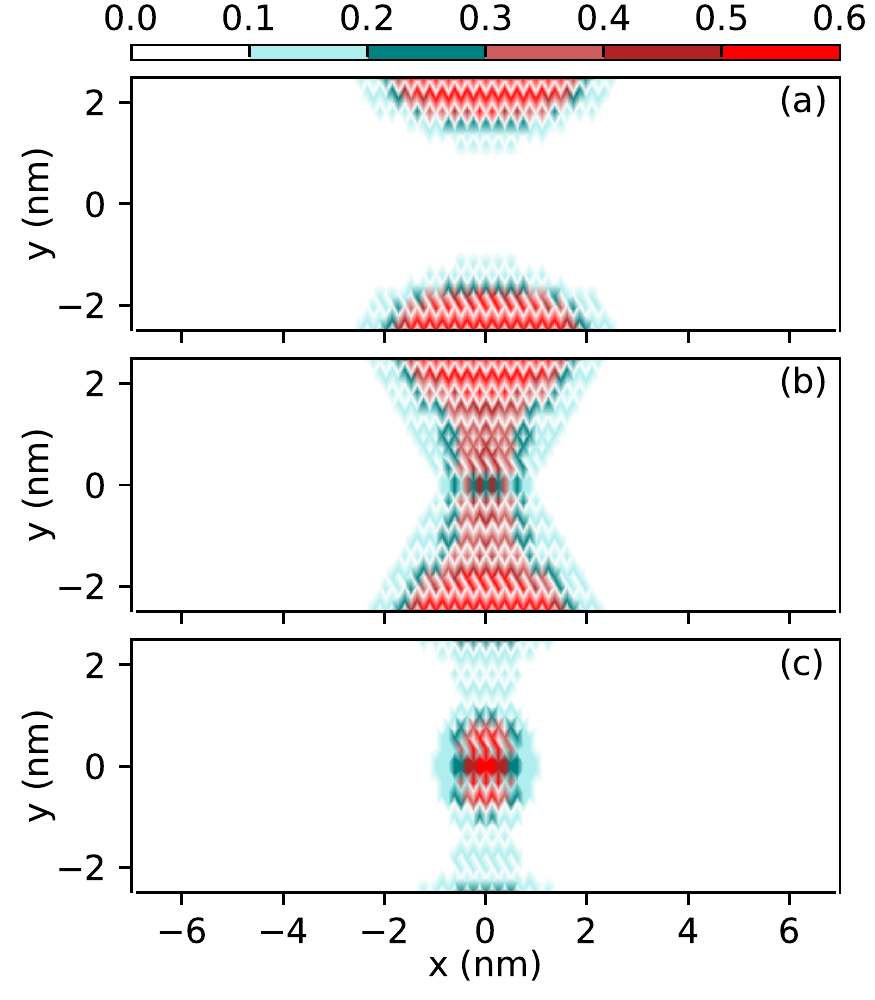}
        \caption{\label{fig:fig12} Contour plot of spatial LDOS of the lowest state in Fig. 10 for three values of the charge: (a) $\beta$ = 0.781 ($E$ = -0.189 eV), (b) $\beta$ = 1.607 ($E$ = -0.432 eV) and (c) $\beta$ = 2.59 ($E$ = -0.871 eV).}
    \end{figure}
    
In Fig.~\ref{fig:fig11}(a) and~\ref{fig:fig11}(b) two cuts of the LDOS of Figs.~\ref{fig:fig10} are shown for $\beta = 1.45$ and $\beta = 2.25$ showing the situation before and after the first bound state that started at $E \approx 0.5$ eV crosses the edge states. In Fig.~\ref{fig:fig11}(a) the series of edge states can be clearly seen as a distinct number of peaks between $E \approx -0.1$ eV and $E \approx -0.4$ eV. At $E \approx -0.1$ eV a very sharp peak can be observed corresponding to the lowest bound state originating from $E \approx 0.5$ eV in Fig.~\ref{fig:fig10}. When the charge is increased to $\beta = 2.25$ as shown in Fig.~\ref{fig:fig10} the low energy bound state of Fig.~\ref{fig:fig11}(a) has turned into a resonant state which is being modulated by the edge states. This behavior is seen in Fig.~\ref{fig:fig11}(b) around $E \approx$ -0.5 eV where the resonant peak shows subpeaks corresponding to the edge states. The interesting behavior of the edge states modulating the resonant states should be a clear signature to look for in experiments.

Now we look into the evolution of the edge states as function of the impurity charge $\beta$. In Fig.~\ref{fig:fig12} we plot the spatial LDOS for the lowest edge state for three values of the impurity charge $\beta$. For small value of the impurity charge (see Fig.~\ref{fig:fig12}(a)) the spatial LDOS is localised at the edges confirming its edge state nature. Interestingly, the edge state turns into an impurity bulk state with increasing charge $\beta$ (see Figs.~\ref{fig:fig12}(b) and~\ref{fig:fig12}(c)). This transition from edge to impurity state explains the change in $\beta$ -dependence (linear versus nonlinear) observed in Fig.~\ref{fig:fig10} and is a clear signature to look for in future experiments. 

    \begin{figure}[htb!]
        \centering
        \includegraphics[width=0.75\columnwidth]{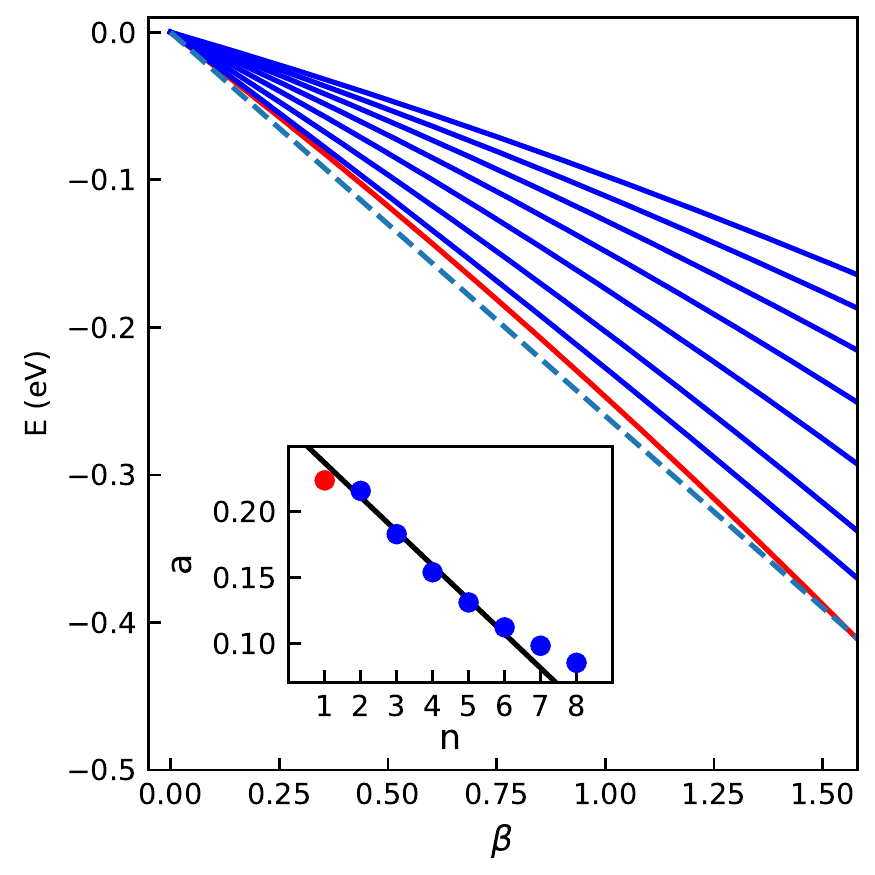}
        \caption{\label{fig:fig13} First 8 states originating from the band of edge states as function of the charge strength $\beta$. The lowest state clearly visible in Fig. 10 is shown in red. The dashed line is the values of the Coulomb potential at the edge. For small $\beta$ the energy states behave as $E = -a\beta$. In the inset the value of the fitting parameter $a$ (in unit of eV) is shown for the different curves. The straight line is given by $a = 0.263 -0.026n$.}
     \end{figure}
     \begin{figure}[htb!]
        \centering
        \includegraphics[width=0.75\columnwidth]{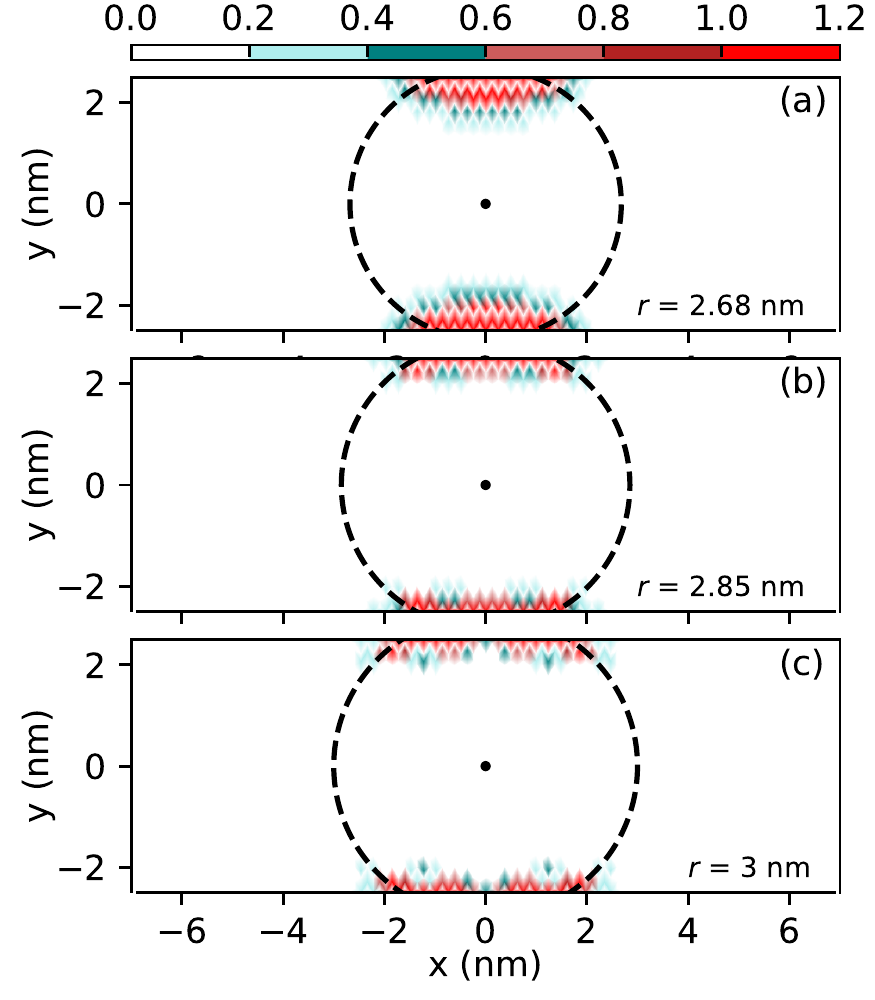}
        \caption{\label{fig:fig14} Contour plot of spatial LDOS calculated for $\beta = 0.8$ for the three lowest states seen in Fig. 13: (a) $E = -0.193$ eV, (b) $E =$ -0.182 eV and (c) $E =$ -0.172 eV. The solid dot shows the position of the charged impurity and the dashed circle indicates the radius(i) of the Coulomb potential at this energy.}
     \end{figure}
    
In Fig.~\ref{fig:fig13} the energy of the fan of bound edge states as function of the impurity charge $\beta$, as derived from the LDOS, is shown for the first 8 impurity edge states. The lowest state which shows a clear edge to impurity state transition discussed earlier is shown in red. The energy levels could be fitted (for the region $0 < \beta < 1$) to $E = -(a \beta + b \beta^2 )$ with \{a, b\} respectively \{0.161, 0.0698\}, \{0.216, 0.012\}, \{0.184, 0.019\}, \{0.154, 0.019\}, \{0.13, 0.018\}, \{0.112, 0.015\}, \{0.098, 0.013\}, and \{0.085, 0.011\} eV. It is evident that for small $\beta$ the energy is linear in $\beta$. This is in contrast with the states of the 2D hydrogen atom which exhibits a quadratic dependence $E \approx \beta^2$~\cite{ref34} while the lowest atomic collapse states in bulk graphene (see Fig.~\ref{fig:fig1}) behaves as $E = (-\beta \hbar v_F/r_0 )exp(-\pi/ \sqrt{\beta ^2 - 0.25})$~\cite{ref8}. Using first order perturbation theory with respect to the Coulomb potential explains the linear $\beta$-dependence of the bound edge states. The quadratic term gives a very small correction to the linear behavior. The drop in $b$ from the first to the second state is a consequence of the fact that the lowest state is turning into a bulk impurity state for large $\beta$. 

In the inset of Fig.~\ref{fig:fig13} we show the fitting parameter $a$ as function of the number of the edge state which shows a linear dependence $a$ (eV) = 0.263 - 0.026$n$ for $1<n<7$. This behavior can be understood from the fact that for small $\beta$ the edge states remain confined at the edge, feeling a broader almost quadratic-like potential for small energy and distances. Consequently, these edge states feel a softer potential, explaining the weaker $\beta$-dependence as function of the edge state number.

\begin{figure}[htb!]
        \centering
        \includegraphics[width=0.75\columnwidth]{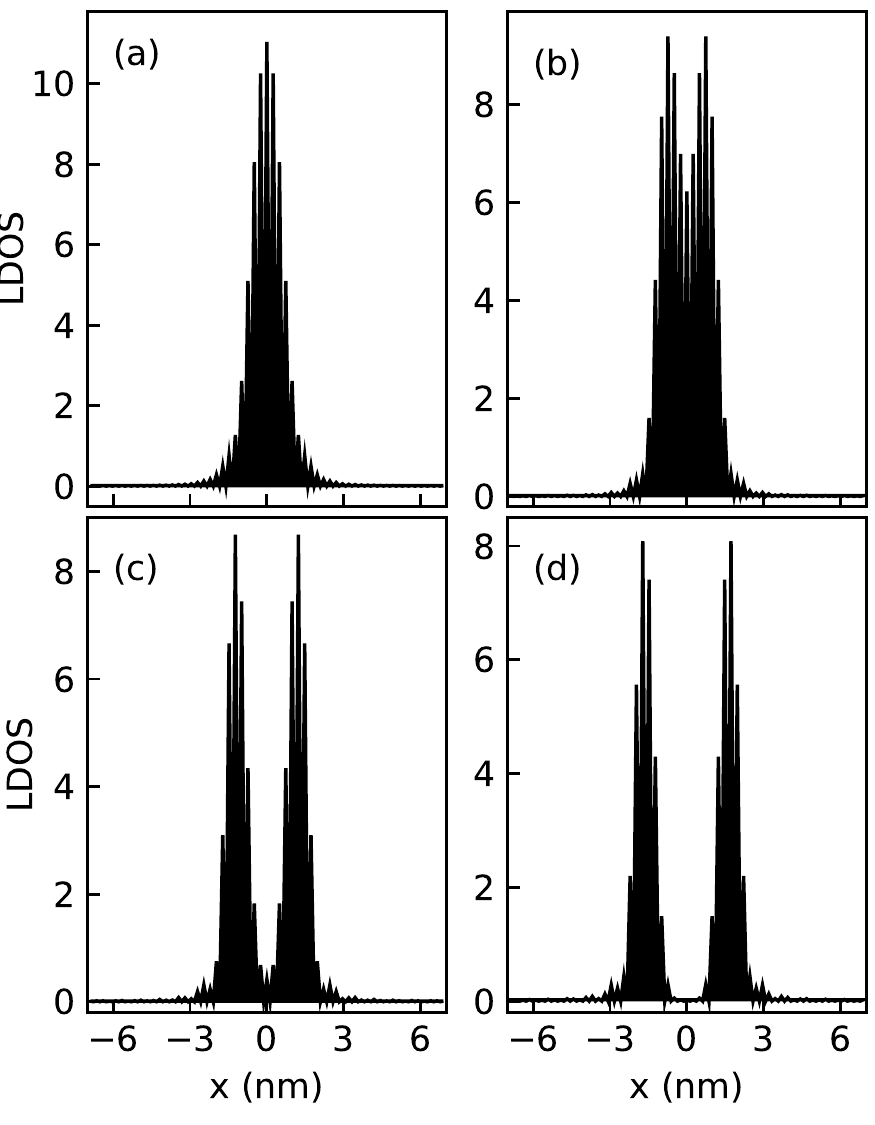}
        \caption{\label{fig:fig15} Cut of the spatial LDOS along the edge, calculated for $\beta = 0.8$ for the four lowest states seen in Fig. 13: (a) $E =$ -0.193 eV, (b) $E =$ -0.182 eV, (c) $E =$ -0.172 eV and (d) $E =$ -0.160 eV.}
     \end{figure}

In Fig.~\ref{fig:fig14} we plot the spatial LDOS for $\beta = 0.8$ for the three lowest states shown in Fig.~\ref{fig:fig13}. All these three states are in the region where they show almost perfect linear behavior as function of the charge $\beta$. We observe that with increasing energy the spatial LDOS localization shifts further away from the center of the nanoribbon (where the Coulomb charge is placed). As a consequence these states feel a weaker shift due to the decay of the Coulomb potential, explaining why they are higher in energy. In Fig.~\ref{fig:fig15} a cut of the spatial LDOS along one of the nanoribbon edges are plotted for the first four states seen in Fig.~\ref{fig:fig13}. The cut is taken along the edge of the nanoribbon. The lowest state consists of one peak while the excited states consist of two peaks which move further away from each other for higher energy states. These figures are somewhat similar but not the same to the electron probability of states found in Ref.~\cite{ref26} for the Coulomb problem in gapped Dirac materials. The two peak structure in LDOS symmetric around $x=0$ can be understood as follows. Lets consider the 1D edge states and take the extreme limit of a flat band. The kinetic energy is quenched and the Dirac equation is reduced to 
    \begin{equation}
        V(x,y_0)\psi(x,y_0)=E\psi(x,y_0)
    \end{equation}where $y_0$ is the position of the edge. This equation has as solution $\psi(x,y_0)\approx\delta(x-x_i)$ where $x_i$ is determined by $V(x_i,y_0)=E$ as shown by the dashed circles in Fig.~\ref{fig:fig14}. Because the Coulomb potential $V(x,y)$ is symmetric around $x=0$ this gives two solutions $x_i=\pm\lvert{x_i}\rvert$ and thus the wave function becomes
    \begin{equation}
        \psi(x,y_0)=c(\delta(x-\lvert{x_i}\rvert)+\delta(x+\lvert{x_i}\rvert)
    \end{equation}
The separation between those $\delta$-functions increase with energy which agrees with Fig.~\ref{fig:fig15}. Those $\delta$-peaks are broadened in our numerical results because the edge states exhibit some small dispersion and the edge states penetrate into the bulk of the nanoribbon exponential decreasing away from the edge.
    
     \begin{figure}[htb!]
        \centering
        \includegraphics[width=0.75\columnwidth]{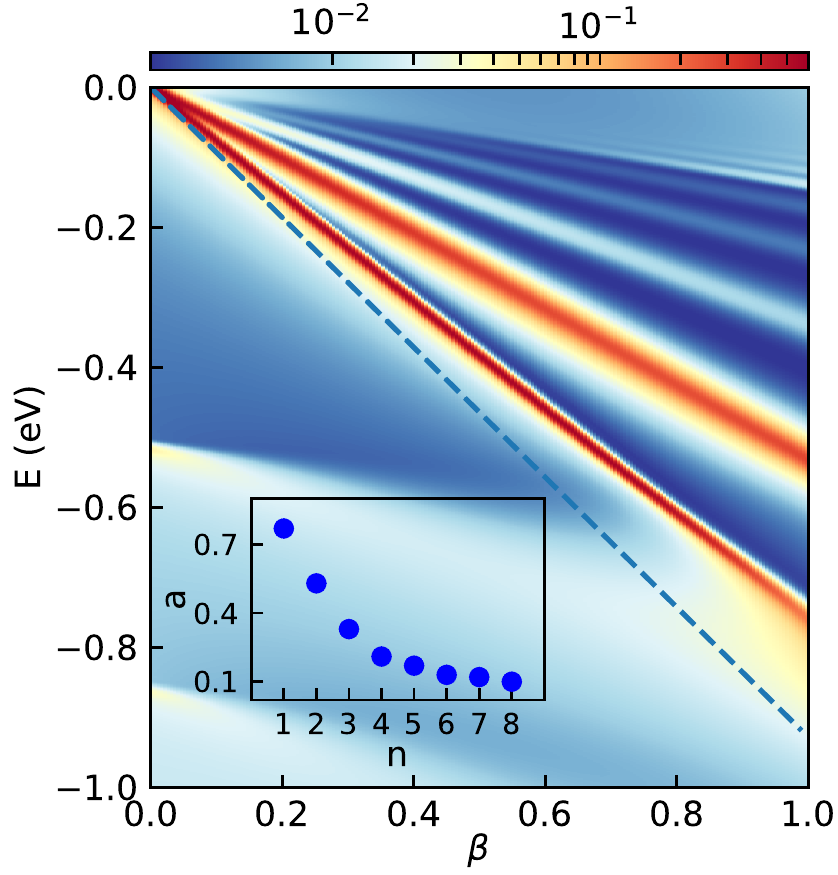}
        \caption{\label{fig:fig16} LDOS calculation for the same nanoribbon as in Fig. 10 but now with a charge which is placed 2 nm from the center of the nanoribbon. The dashed line is the value of the Coulomb potential at the closest edge. In the inset we show the fitting parameter $a$ (in units of eV) as function of the number of the edge state for the five lowest states.}
    \end{figure}
    \begin{figure}[htb!]
        \centering
        \includegraphics[width=0.75\columnwidth]{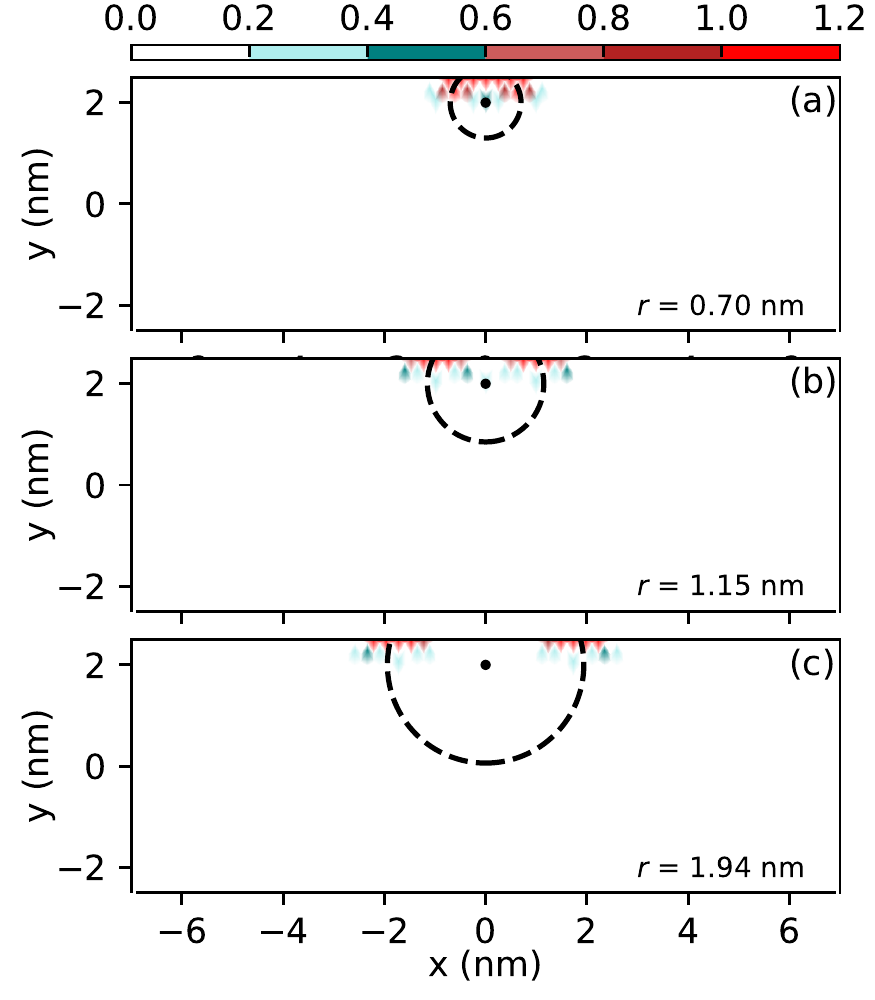}
        \caption{\label{fig:fig17} Contour plot of spatial LDOS calculated for $\beta = 0.54$ for three lowest states seen in Fig. 16: (a) $E = -0.414$ eV, (b) $E = -0.284$ eV and (c) $E = -0.177$ eV. The solid dot shows the position of the charged impurity and the dashed circle indicates the radius(i) of the Coulomb potential at this energy.}
    \end{figure} 
    
From Fig.~\ref{fig:fig10} and the previous discussion it is clear that the states originating from the edge states are narrow and consequently represent bound states. However, at first sight this seems strange because looking at Fig.~\ref{fig:fig9} reveals that the LDOS is not zero below these edge states. Consequently one may expect that these edge states will hybridize with the underlying continuum turning into resonant states (similar to the upper band states previously discussed for armchair nanoribbons). However, this seems not the case. From our previous results for armchair nanoribbons one may expect that this behavior can be related to the symmetric placement of the charge. In order to confirm whether or not this suspicion is correct we calculated the LDOS at the impurity for a charge placed 2 nm from the center of the nanoribbon in Fig.~\ref{fig:fig16}. We limited the figure to the edge states since they are here of interest. Because the states are more clearly separated it is now even more clear how these states originate from $E = 0$ eV. We checked that the width of those peaks in the LDOS scales with the numerical broadening confirming that they are bound states. For $\beta<1$ the energy of these states is linear the charge $\beta$ similar to the states shown in Fig.~\ref{fig:fig10}. This time the states can be almost perfectly fitted to $E =$ $-a\beta$ with \{$a$\} respectively \{0.77\}, \{0.53\}, \{0.33\}, \{0.21\}, \{0.17\}, \{0.13\}, \{0.17\},\{0.12\}, and \{0.10\} eV for the eight lowest states seen in Fig.~\ref{fig:fig16}. In the inset of Fig.~\ref{fig:fig16} the fitting parameter $a$ is shown as function of the number of the edge state. In contrast with the results for a symmetrically placed charge shown in the inset of Fig.~\ref{fig:fig13} no linear dependence is observed and the states depend more strongly on the impurity charge $\beta$. The reason is that the Coulomb charge is placed much closer to the edge, consequently the edge states feel a much deeper Coulomb potential instead of the broader potential felt by the edge states discussed earlier. Therefore the edge states are more strongly influenced by the charge explaining their more profound dependence on the charge $\beta$.

In Fig.~\ref{fig:fig17} the spatial LDOS is shown for $\beta = 0.54$ for the three lowest edge states seen in Fig.~\ref{fig:fig16}. The behavior of these states is very similar to the ones shown in Fig.~\ref{fig:fig14} for the symmetrically placed charge with the difference that the state is now entirely localized at only one edge, i.e. the edge closest to the potential center.

\section{Conclusion}\label{sec:6}
    
We investigated how the finite width of graphene nanoribbons influences the atomic collapse phenomenon and found very different physical behaviours depending on the type of edges.
    
We showed with tight binding calculations that the manifestation of the atomic collapse in graphene nanoribbons is fundamentally different from its manifestation in pristine graphene. In both armchair and zig-zag nanoribbons bound states turn into atomic collapse states when entering the lower continuum. This kind of behavior mimics closely the predicted atomic collapse in relativistic physics: bound states in the mass gap turn into quasi-bound states when entering the negative continuum. Therefore, the experimental study of atomic collapse in graphene nanoribbons could pave the way to the first observation of the true analog of the relativistic atomic collapse effect.

We showed that in the case of zig-zag nanoribbons the well known edge states lead to a modulation of the quasi-bound states when they cross the band of edge states. This modulation should be measurable in experiments when probing with an STM tip.
    
Furthermore we showed that the atomic collapse in graphene nanoribbons differs from the manifestation in pristine graphene in the following ways: i) instead of the sudden appearance of quasi-bound states in pristine graphene a gradual bound state to quasi-bound state transition is predicted in nanoribbons providing a very close analog of atomic collapse in relativistic physics. ii) The appearance of multiple energy bands leads to a richer spectrum as compared to pristine graphene with the appearance of multiple quasi-bound electron states below each energy band. iii) In the case of zig-zag nanoribbons an extra band of bound states appear, which are predominantly localized at the edge of the nanoribbon. The flat band character of those edge states are the origin of these quasi 1D bound states and its LDOS consists of two identical peaks whose separation increases with energy. The bound character of these states are also a consequence of the fact that scattering on a zig-zag edge does not allow intervalley scattering. 

Last we would like to make a remark about the experimental feasibility of observing the effects predicted in this paper. The production of different types of nanoribbons have been realized over the last few years~\cite{ref35,ref36,ref37,ref38,ref39}. Recently~\cite{ref40}, perfect edges in 2D materials was realized by using a combination of top-down lithography with a near anisotropic wet etching process. Placing the charges on the nanoribbons should be possible using an STM tip as realized in Ref.~\cite{ref9} for bulk graphene. Such an STM tip can also be used to measure the LDOS. Alternatively, it may be possible to mimic the potentials created by charged dimers using an STM tip~\cite{ref11} or a charged vacancy~\cite{ref10} producing similar effects. We hope that given the number of recent experiments it should be possible to test our predictions in the near future.

\begin{acknowledgments}
    We thank Matthias Van der Donck for fruitful discussions. This work was supported by the National Natural Science Foundation of China (Grant Nos. 62004053, 61874038 and 61704040), National Key R\&D Program Grant 2018YFE0120000, the scholarship from China Scholarship Council (CSC: 201908330548) and Research Foundation of Flanders (FWO-V1) through an aspirant research grant for RVP and a postdoc fellowship for BVD.
\end{acknowledgments}

\bibliography{bibliography}

\begin{thebibliography}{40}%
\makeatletter
\providecommand \@ifxundefined [1]{%
 \@ifx{#1\undefined}
}%
\providecommand \@ifnum [1]{%
 \ifnum #1\expandafter \@firstoftwo
 \else \expandafter \@secondoftwo
 \fi
}%
\providecommand \@ifx [1]{%
 \ifx #1\expandafter \@firstoftwo
 \else \expandafter \@secondoftwo
 \fi
}%
\providecommand \natexlab [1]{#1}%
\providecommand \enquote  [1]{``#1''}%
\providecommand \bibnamefont  [1]{#1}%
\providecommand \bibfnamefont [1]{#1}%
\providecommand \citenamefont [1]{#1}%
\providecommand \href@noop [0]{\@secondoftwo}%
\providecommand \href [0]{\begingroup \@sanitize@url \@href}%
\providecommand \@href[1]{\@@startlink{#1}\@@href}%
\providecommand \@@href[1]{\endgroup#1\@@endlink}%
\providecommand \@sanitize@url [0]{\catcode `\\12\catcode `\$12\catcode
  `\&12\catcode `\#12\catcode `\^12\catcode `\_12\catcode `\%12\relax}%
\providecommand \@@startlink[1]{}%
\providecommand \@@endlink[0]{}%
\providecommand \url  [0]{\begingroup\@sanitize@url \@url }%
\providecommand \@url [1]{\endgroup\@href {#1}{\urlprefix }}%
\providecommand \urlprefix  [0]{URL }%
\providecommand \Eprint [0]{\href }%
\providecommand \doibase [0]{https://doi.org/}%
\providecommand \selectlanguage [0]{\@gobble}%
\providecommand \bibinfo  [0]{\@secondoftwo}%
\providecommand \bibfield  [0]{\@secondoftwo}%
\providecommand \translation [1]{[#1]}%
\providecommand \BibitemOpen [0]{}%
\providecommand \bibitemStop [0]{}%
\providecommand \bibitemNoStop [0]{.\EOS\space}%
\providecommand \EOS [0]{\spacefactor3000\relax}%
\providecommand \BibitemShut  [1]{\csname bibitem#1\endcsname}%
\let\auto@bib@innerbib\@empty
\bibitem [{\citenamefont {Greiner}\ \emph {et~al.}(1985)\citenamefont
  {Greiner}, \citenamefont {M\"{u}ller},\ and\ \citenamefont
  {Rafelski}}]{ref1}%
  \BibitemOpen
  \bibfield  {author} {\bibinfo {author} {\bibfnamefont {W.}~\bibnamefont
  {Greiner}}, \bibinfo {author} {\bibfnamefont {B.}~\bibnamefont
  {M\"{u}ller}},\ and\ \bibinfo {author} {\bibfnamefont {J.}~\bibnamefont
  {Rafelski}},\ }\href {https://doi.org/10.1007/978-3-642-82272-8} {\emph
  {\bibinfo {title} {Quantum Electrodynamics of Strong Fields}}}\ (\bibinfo
  {publisher} {Springer Berlin Heidelberg},\ \bibinfo {year}
  {1985})\BibitemShut {NoStop}%
\bibitem [{\citenamefont {Pomeranchuk}\ and\ \citenamefont
  {Smorodinsky}(1945)}]{ref2}%
  \BibitemOpen
  \bibfield  {author} {\bibinfo {author} {\bibfnamefont {I.}~\bibnamefont
  {Pomeranchuk}}\ and\ \bibinfo {author} {\bibfnamefont {Y.}~\bibnamefont
  {Smorodinsky}},\ }\href@noop {} {\bibfield  {journal} {\bibinfo  {journal}
  {J. Phys. USSR}\ }\textbf {\bibinfo {volume} {9}},\ \bibinfo {pages} {97}
  (\bibinfo {year} {1945})}\BibitemShut {NoStop}%
\bibitem [{\citenamefont {Zeldovich}\ and\ \citenamefont {Popov}(1972)}]{ref3}%
  \BibitemOpen
  \bibfield  {author} {\bibinfo {author} {\bibfnamefont {Y.~B.}\ \bibnamefont
  {Zeldovich}}\ and\ \bibinfo {author} {\bibfnamefont {V.~S.}\ \bibnamefont
  {Popov}},\ }\href@noop {} {\bibfield  {journal} {\bibinfo  {journal} {Sov.
  Phys. Usp.}\ }\textbf {\bibinfo {volume} {14}},\ \bibinfo {pages} {16}
  (\bibinfo {year} {1972})}\BibitemShut {NoStop}%
\bibitem [{\citenamefont {Shytov}\ \emph {et~al.}(2009)\citenamefont {Shytov},
  \citenamefont {Rudner}, \citenamefont {Gu}, \citenamefont {Katsnelson},\ and\
  \citenamefont {Levitov}}]{ref4}%
  \BibitemOpen
  \bibfield  {author} {\bibinfo {author} {\bibfnamefont {A.}~\bibnamefont
  {Shytov}}, \bibinfo {author} {\bibfnamefont {M.}~\bibnamefont {Rudner}},
  \bibinfo {author} {\bibfnamefont {N.}~\bibnamefont {Gu}}, \bibinfo {author}
  {\bibfnamefont {M.}~\bibnamefont {Katsnelson}},\ and\ \bibinfo {author}
  {\bibfnamefont {L.}~\bibnamefont {Levitov}},\ }\href
  {https://doi.org/10.1038/nature04233} {\bibfield  {journal} {\bibinfo
  {journal} {Sol. Stat. Comm.}\ }\textbf {\bibinfo {volume} {149}},\ \bibinfo
  {pages} {1087} (\bibinfo {year} {2009})}\BibitemShut {NoStop}%
\bibitem [{\citenamefont {Schweppe}\ \emph {et~al.}(1983)\citenamefont
  {Schweppe}, \citenamefont {Gruppe}, \citenamefont {Bethge}, \citenamefont
  {Bokemeyer}, \citenamefont {Cowan}, \citenamefont {Folger}, \citenamefont
  {Greenberg}, \citenamefont {Grein}, \citenamefont {Ito}, \citenamefont
  {Schule}, \citenamefont {Schwalm}, \citenamefont {Stiebing}, \citenamefont
  {Trautmann}, \citenamefont {Vincent},\ and\ \citenamefont
  {Waldschmidt}}]{ref5}%
  \BibitemOpen
  \bibfield  {author} {\bibinfo {author} {\bibfnamefont {J.}~\bibnamefont
  {Schweppe}}, \bibinfo {author} {\bibfnamefont {A.}~\bibnamefont {Gruppe}},
  \bibinfo {author} {\bibfnamefont {K.}~\bibnamefont {Bethge}}, \bibinfo
  {author} {\bibfnamefont {H.}~\bibnamefont {Bokemeyer}}, \bibinfo {author}
  {\bibfnamefont {T.}~\bibnamefont {Cowan}}, \bibinfo {author} {\bibfnamefont
  {H.}~\bibnamefont {Folger}}, \bibinfo {author} {\bibfnamefont {J.~S.}\
  \bibnamefont {Greenberg}}, \bibinfo {author} {\bibfnamefont {H.}~\bibnamefont
  {Grein}}, \bibinfo {author} {\bibfnamefont {S.}~\bibnamefont {Ito}}, \bibinfo
  {author} {\bibfnamefont {R.}~\bibnamefont {Schule}}, \bibinfo {author}
  {\bibfnamefont {D.}~\bibnamefont {Schwalm}}, \bibinfo {author} {\bibfnamefont
  {K.~E.}\ \bibnamefont {Stiebing}}, \bibinfo {author} {\bibfnamefont
  {N.}~\bibnamefont {Trautmann}}, \bibinfo {author} {\bibfnamefont
  {P.}~\bibnamefont {Vincent}},\ and\ \bibinfo {author} {\bibfnamefont
  {M.}~\bibnamefont {Waldschmidt}},\ }\href
  {https://doi.org/10.1103/PhysRevLett.51.2261} {\bibfield  {journal} {\bibinfo
   {journal} {Phys. Rev. Lett.}\ }\textbf {\bibinfo {volume} {51}},\ \bibinfo
  {pages} {2261} (\bibinfo {year} {1983})}\BibitemShut {NoStop}%
\bibitem [{\citenamefont {Cowan}\ \emph {et~al.}(1985)\citenamefont {Cowan},
  \citenamefont {Backe}, \citenamefont {Begemann}, \citenamefont {Bethge},
  \citenamefont {Bokemeyer}, \citenamefont {Folger}, \citenamefont {Greenberg},
  \citenamefont {Grein}, \citenamefont {Gruppe}, \citenamefont {Kido},
  \citenamefont {Klver}, \citenamefont {Schwalm}, \citenamefont {Schweppe},
  \citenamefont {Stiebing}, \citenamefont {Trautmann},\ and\ \citenamefont
  {Vincent}}]{ref6}%
  \BibitemOpen
  \bibfield  {author} {\bibinfo {author} {\bibfnamefont {T.}~\bibnamefont
  {Cowan}}, \bibinfo {author} {\bibfnamefont {H.}~\bibnamefont {Backe}},
  \bibinfo {author} {\bibfnamefont {M.}~\bibnamefont {Begemann}}, \bibinfo
  {author} {\bibfnamefont {K.}~\bibnamefont {Bethge}}, \bibinfo {author}
  {\bibfnamefont {H.}~\bibnamefont {Bokemeyer}}, \bibinfo {author}
  {\bibfnamefont {H.}~\bibnamefont {Folger}}, \bibinfo {author} {\bibfnamefont
  {J.~S.}\ \bibnamefont {Greenberg}}, \bibinfo {author} {\bibfnamefont
  {H.}~\bibnamefont {Grein}}, \bibinfo {author} {\bibfnamefont
  {A.}~\bibnamefont {Gruppe}}, \bibinfo {author} {\bibfnamefont
  {Y.}~\bibnamefont {Kido}}, \bibinfo {author} {\bibfnamefont {M.}~\bibnamefont
  {Klver}}, \bibinfo {author} {\bibfnamefont {D.}~\bibnamefont {Schwalm}},
  \bibinfo {author} {\bibfnamefont {J.}~\bibnamefont {Schweppe}}, \bibinfo
  {author} {\bibfnamefont {K.~E.}\ \bibnamefont {Stiebing}}, \bibinfo {author}
  {\bibfnamefont {N.}~\bibnamefont {Trautmann}},\ and\ \bibinfo {author}
  {\bibfnamefont {P.}~\bibnamefont {Vincent}},\ }\href
  {https://doi.org/10.1103/PhysRevLett.54.1761} {\bibfield  {journal} {\bibinfo
   {journal} {Phys. Rev. Lett.}\ }\textbf {\bibinfo {volume} {54}},\ \bibinfo
  {pages} {1761} (\bibinfo {year} {1985})}\BibitemShut {NoStop}%
\bibitem [{\citenamefont {Pereira}\ \emph {et~al.}(2007)\citenamefont
  {Pereira}, \citenamefont {Nilsson},\ and\ \citenamefont
  {Castro~Neto}}]{ref7}%
  \BibitemOpen
  \bibfield  {author} {\bibinfo {author} {\bibfnamefont {V.~M.}\ \bibnamefont
  {Pereira}}, \bibinfo {author} {\bibfnamefont {J.}~\bibnamefont {Nilsson}},\
  and\ \bibinfo {author} {\bibfnamefont {A.~H.}\ \bibnamefont {Castro~Neto}},\
  }\href {https://doi.org/10.1103/PhysRevLett.99.166802} {\bibfield  {journal}
  {\bibinfo  {journal} {Phys. Rev. Lett.}\ }\textbf {\bibinfo {volume} {99}},\
  \bibinfo {pages} {166802} (\bibinfo {year} {2007})}\BibitemShut {NoStop}%
\bibitem [{\citenamefont {Shytov}\ \emph {et~al.}(2007)\citenamefont {Shytov},
  \citenamefont {Katsnelson},\ and\ \citenamefont {Levitov}}]{ref8}%
  \BibitemOpen
  \bibfield  {author} {\bibinfo {author} {\bibfnamefont {A.~V.}\ \bibnamefont
  {Shytov}}, \bibinfo {author} {\bibfnamefont {M.~I.}\ \bibnamefont
  {Katsnelson}},\ and\ \bibinfo {author} {\bibfnamefont {L.~S.}\ \bibnamefont
  {Levitov}},\ }\href {https://doi.org/10.1103/PhysRevLett.99.236801}
  {\bibfield  {journal} {\bibinfo  {journal} {Phys. Rev. Lett.}\ }\textbf
  {\bibinfo {volume} {99}},\ \bibinfo {pages} {236801} (\bibinfo {year}
  {2007})}\BibitemShut {NoStop}%
\bibitem [{\citenamefont {Wang}\ \emph {et~al.}(2013)\citenamefont {Wang},
  \citenamefont {Wong}, \citenamefont {Shytov}, \citenamefont {Brar},
  \citenamefont {Choi}, \citenamefont {Wu}, \citenamefont {Tsai}, \citenamefont
  {Regan}, \citenamefont {Zettl}, \citenamefont {Kawakami}, \citenamefont
  {Louie}, \citenamefont {Levitov},\ and\ \citenamefont {Crommie}}]{ref9}%
  \BibitemOpen
  \bibfield  {author} {\bibinfo {author} {\bibfnamefont {Y.}~\bibnamefont
  {Wang}}, \bibinfo {author} {\bibfnamefont {D.}~\bibnamefont {Wong}}, \bibinfo
  {author} {\bibfnamefont {A.~V.}\ \bibnamefont {Shytov}}, \bibinfo {author}
  {\bibfnamefont {V.~W.}\ \bibnamefont {Brar}}, \bibinfo {author}
  {\bibfnamefont {S.}~\bibnamefont {Choi}}, \bibinfo {author} {\bibfnamefont
  {Q.}~\bibnamefont {Wu}}, \bibinfo {author} {\bibfnamefont {H.-Z.}\
  \bibnamefont {Tsai}}, \bibinfo {author} {\bibfnamefont {W.}~\bibnamefont
  {Regan}}, \bibinfo {author} {\bibfnamefont {A.}~\bibnamefont {Zettl}},
  \bibinfo {author} {\bibfnamefont {R.~K.}\ \bibnamefont {Kawakami}}, \bibinfo
  {author} {\bibfnamefont {S.~G.}\ \bibnamefont {Louie}}, \bibinfo {author}
  {\bibfnamefont {L.~S.}\ \bibnamefont {Levitov}},\ and\ \bibinfo {author}
  {\bibfnamefont {M.~F.}\ \bibnamefont {Crommie}},\ }\href
  {https://doi.org/10.1126/science.1234320} {\bibfield  {journal} {\bibinfo
  {journal} {Science}\ }\textbf {\bibinfo {volume} {340}},\ \bibinfo {pages}
  {734} (\bibinfo {year} {2013})}\BibitemShut {NoStop}%
\bibitem [{\citenamefont {Mao}\ \emph {et~al.}(2016)\citenamefont {Mao},
  \citenamefont {Jiang}, \citenamefont {Moldovan}, \citenamefont {Li},
  \citenamefont {Watanabe}, \citenamefont {Taniguchi}, \citenamefont {Masir},
  \citenamefont {Peeters},\ and\ \citenamefont {Andrei}}]{ref10}%
  \BibitemOpen
  \bibfield  {author} {\bibinfo {author} {\bibfnamefont {J.~H.}\ \bibnamefont
  {Mao}}, \bibinfo {author} {\bibfnamefont {Y.~H.}\ \bibnamefont {Jiang}},
  \bibinfo {author} {\bibfnamefont {D.}~\bibnamefont {Moldovan}}, \bibinfo
  {author} {\bibfnamefont {G.~H.}\ \bibnamefont {Li}}, \bibinfo {author}
  {\bibfnamefont {K.}~\bibnamefont {Watanabe}}, \bibinfo {author}
  {\bibfnamefont {T.}~\bibnamefont {Taniguchi}}, \bibinfo {author}
  {\bibfnamefont {M.~R.}\ \bibnamefont {Masir}}, \bibinfo {author}
  {\bibfnamefont {F.~M.}\ \bibnamefont {Peeters}},\ and\ \bibinfo {author}
  {\bibfnamefont {E.~Y.}\ \bibnamefont {Andrei}},\ }\href
  {https://doi.org/10.1038/nphys3665} {\bibfield  {journal} {\bibinfo
  {journal} {Nat. Phys.}\ }\textbf {\bibinfo {volume} {12}},\ \bibinfo {pages}
  {545} (\bibinfo {year} {2016})}\BibitemShut {NoStop}%
\bibitem [{\citenamefont {Jiang}\ \emph {et~al.}(2017)\citenamefont {Jiang},
  \citenamefont {Mao}, \citenamefont {Moldovan}, \citenamefont {Masir},
  \citenamefont {Li}, \citenamefont {Watanabe}, \citenamefont {Taniguchi},
  \citenamefont {Peeters},\ and\ \citenamefont {Andrei}}]{ref11}%
  \BibitemOpen
  \bibfield  {author} {\bibinfo {author} {\bibfnamefont {Y.~H.}\ \bibnamefont
  {Jiang}}, \bibinfo {author} {\bibfnamefont {J.~H.}\ \bibnamefont {Mao}},
  \bibinfo {author} {\bibfnamefont {D.}~\bibnamefont {Moldovan}}, \bibinfo
  {author} {\bibfnamefont {M.~R.}\ \bibnamefont {Masir}}, \bibinfo {author}
  {\bibfnamefont {G.~H.}\ \bibnamefont {Li}}, \bibinfo {author} {\bibfnamefont
  {K.}~\bibnamefont {Watanabe}}, \bibinfo {author} {\bibfnamefont
  {T.}~\bibnamefont {Taniguchi}}, \bibinfo {author} {\bibfnamefont {F.~M.}\
  \bibnamefont {Peeters}},\ and\ \bibinfo {author} {\bibfnamefont {E.~Y.}\
  \bibnamefont {Andrei}},\ }\href {https://doi.org/10.1038/nnano.2017.181}
  {\bibfield  {journal} {\bibinfo  {journal} {Nat. Nanotech.}\ }\textbf
  {\bibinfo {volume} {12}},\ \bibinfo {pages} {1045} (\bibinfo {year}
  {2017})}\BibitemShut {NoStop}%
\bibitem [{\citenamefont {Lu}\ \emph {et~al.}(2019)\citenamefont {Lu},
  \citenamefont {Tsai}, \citenamefont {Tatan}, \citenamefont {Wickenburg},
  \citenamefont {Omrani}, \citenamefont {Wong}, \citenamefont {Riss},
  \citenamefont {Piatti}, \citenamefont {Watanabe}, \citenamefont {Taniguchi},
  \citenamefont {Zettl}, \citenamefont {Pereira},\ and\ \citenamefont
  {Crommie}}]{ref12}%
  \BibitemOpen
  \bibfield  {author} {\bibinfo {author} {\bibfnamefont {J.}~\bibnamefont
  {Lu}}, \bibinfo {author} {\bibfnamefont {H.~Z.}\ \bibnamefont {Tsai}},
  \bibinfo {author} {\bibfnamefont {A.~N.}\ \bibnamefont {Tatan}}, \bibinfo
  {author} {\bibfnamefont {S.}~\bibnamefont {Wickenburg}}, \bibinfo {author}
  {\bibfnamefont {A.~A.}\ \bibnamefont {Omrani}}, \bibinfo {author}
  {\bibfnamefont {D.}~\bibnamefont {Wong}}, \bibinfo {author} {\bibfnamefont
  {A.}~\bibnamefont {Riss}}, \bibinfo {author} {\bibfnamefont {E.}~\bibnamefont
  {Piatti}}, \bibinfo {author} {\bibfnamefont {K.}~\bibnamefont {Watanabe}},
  \bibinfo {author} {\bibfnamefont {T.}~\bibnamefont {Taniguchi}}, \bibinfo
  {author} {\bibfnamefont {A.}~\bibnamefont {Zettl}}, \bibinfo {author}
  {\bibfnamefont {V.~M.}\ \bibnamefont {Pereira}},\ and\ \bibinfo {author}
  {\bibfnamefont {M.~F.}\ \bibnamefont {Crommie}},\ }\href
  {https://doi.org/10.1038/s41467-019-08371-2} {\bibfield  {journal} {\bibinfo
  {journal} {Nat. Commun.}\ }\textbf {\bibinfo {volume} {10}},\ \bibinfo
  {pages} {477} (\bibinfo {year} {2019})}\BibitemShut {NoStop}%
\bibitem [{\citenamefont {Klöpfer}\ \emph {et~al.}(2014)\citenamefont
  {Klöpfer}, \citenamefont {Martino}, \citenamefont {Matrasulov},\ and\
  \citenamefont {Egger}}]{ref13}%
  \BibitemOpen
  \bibfield  {author} {\bibinfo {author} {\bibfnamefont {D.}~\bibnamefont
  {Klöpfer}}, \bibinfo {author} {\bibfnamefont {A.~D.}\ \bibnamefont
  {Martino}}, \bibinfo {author} {\bibfnamefont {D.~U.}\ \bibnamefont
  {Matrasulov}},\ and\ \bibinfo {author} {\bibfnamefont {R.}~\bibnamefont
  {Egger}},\ }\href {https://doi.org/10.1140/epjb/e2014-50414-8} {\bibfield
  {journal} {\bibinfo  {journal} {Eur. Phys. J. B}\ }\textbf {\bibinfo {volume}
  {87}},\ \bibinfo {pages} {187} (\bibinfo {year} {2014})}\BibitemShut
  {NoStop}%
\bibitem [{\citenamefont {Gorbar}\ \emph
  {et~al.}(2015{\natexlab{a}})\citenamefont {Gorbar}, \citenamefont {Gusynin},\
  and\ \citenamefont {Sobol}}]{ref14}%
  \BibitemOpen
  \bibfield  {author} {\bibinfo {author} {\bibfnamefont {E.~V.}\ \bibnamefont
  {Gorbar}}, \bibinfo {author} {\bibfnamefont {V.~P.}\ \bibnamefont
  {Gusynin}},\ and\ \bibinfo {author} {\bibfnamefont {O.~O.}\ \bibnamefont
  {Sobol}},\ }\href
  {https://iopscience.iop.org/article/10.1209/0295-5075/111/37003} {\bibfield
  {journal} {\bibinfo  {journal} {Eur. Phys. Lett.}\ }\textbf {\bibinfo
  {volume} {111}},\ \bibinfo {pages} {37003} (\bibinfo {year}
  {2015}{\natexlab{a}})}\BibitemShut {NoStop}%
\bibitem [{\citenamefont {Gorbar}\ \emph
  {et~al.}(2015{\natexlab{b}})\citenamefont {Gorbar}, \citenamefont {Gusynin},\
  and\ \citenamefont {Sobol}}]{ref15}%
  \BibitemOpen
  \bibfield  {author} {\bibinfo {author} {\bibfnamefont {E.~V.}\ \bibnamefont
  {Gorbar}}, \bibinfo {author} {\bibfnamefont {V.~P.}\ \bibnamefont
  {Gusynin}},\ and\ \bibinfo {author} {\bibfnamefont {O.~O.}\ \bibnamefont
  {Sobol}},\ }\href {https://doi.org/10.1103/PhysRevB.92.235417} {\bibfield
  {journal} {\bibinfo  {journal} {Phys. Rev. B}\ }\textbf {\bibinfo {volume}
  {92}},\ \bibinfo {pages} {235417} (\bibinfo {year}
  {2015}{\natexlab{b}})}\BibitemShut {NoStop}%
\bibitem [{\citenamefont {Martino}\ \emph {et~al.}(2014)\citenamefont
  {Martino}, \citenamefont {Klöpfer}, \citenamefont {matrasulov},\ and\
  \citenamefont {Egger}}]{ref16}%
  \BibitemOpen
  \bibfield  {author} {\bibinfo {author} {\bibfnamefont {A.~D.}\ \bibnamefont
  {Martino}}, \bibinfo {author} {\bibfnamefont {D.}~\bibnamefont {Klöpfer}},
  \bibinfo {author} {\bibfnamefont {D.}~\bibnamefont {matrasulov}},\ and\
  \bibinfo {author} {\bibfnamefont {R.}~\bibnamefont {Egger}},\ }\href
  {https://doi.org/10.1103/PhysRevLett.112.186603} {\bibfield  {journal}
  {\bibinfo  {journal} {Phys. Rev. Lett.}\ }\textbf {\bibinfo {volume} {112}},\
  \bibinfo {pages} {186603} (\bibinfo {year} {2014})}\BibitemShut {NoStop}%
\bibitem [{\citenamefont {Pottelberge}\ \emph
  {et~al.}(2018{\natexlab{a}})\citenamefont {Pottelberge}, \citenamefont
  {Duppen},\ and\ \citenamefont {Peeters}}]{ref17}%
  \BibitemOpen
  \bibfield  {author} {\bibinfo {author} {\bibfnamefont {R.~V.}\ \bibnamefont
  {Pottelberge}}, \bibinfo {author} {\bibfnamefont {B.~V.}\ \bibnamefont
  {Duppen}},\ and\ \bibinfo {author} {\bibfnamefont {F.~M.}\ \bibnamefont
  {Peeters}},\ }\href {https://doi.org/10.1103/PhysRevB.98.165420} {\bibfield
  {journal} {\bibinfo  {journal} {Phys. Rev. B}\ }\textbf {\bibinfo {volume}
  {98}},\ \bibinfo {pages} {165420} (\bibinfo {year}
  {2018}{\natexlab{a}})}\BibitemShut {NoStop}%
\bibitem [{\citenamefont {Sobol}\ \emph {et~al.}(2013)\citenamefont {Sobol},
  \citenamefont {Gorbar},\ and\ \citenamefont {Gusynin}}]{ref18}%
  \BibitemOpen
  \bibfield  {author} {\bibinfo {author} {\bibfnamefont {O.~O.}\ \bibnamefont
  {Sobol}}, \bibinfo {author} {\bibfnamefont {E.~V.}\ \bibnamefont {Gorbar}},\
  and\ \bibinfo {author} {\bibfnamefont {V.~P.}\ \bibnamefont {Gusynin}},\
  }\href {https://doi.org/10.1103/PhysRevB.88.205116} {\bibfield  {journal}
  {\bibinfo  {journal} {Phys. Rev. B}\ }\textbf {\bibinfo {volume} {88}},\
  \bibinfo {pages} {205116} (\bibinfo {year} {2013})}\BibitemShut {NoStop}%
\bibitem [{\citenamefont {Pottelberge}\ \emph {et~al.}(2019)\citenamefont
  {Pottelberge}, \citenamefont {Moldovan}, \citenamefont {Milovanovi{\'{c}}},\
  and\ \citenamefont {Peeters}}]{ref19}%
  \BibitemOpen
  \bibfield  {author} {\bibinfo {author} {\bibfnamefont {R.~V.}\ \bibnamefont
  {Pottelberge}}, \bibinfo {author} {\bibfnamefont {D.}~\bibnamefont
  {Moldovan}}, \bibinfo {author} {\bibfnamefont {S.~P.}\ \bibnamefont
  {Milovanovi{\'{c}}}},\ and\ \bibinfo {author} {\bibfnamefont {F.~M.}\
  \bibnamefont {Peeters}},\ }\href {https://doi.org/10.1088/2053-1583/ab3feb}
  {\bibfield  {journal} {\bibinfo  {journal} {2D Mater.}\ }\textbf {\bibinfo
  {volume} {6}},\ \bibinfo {pages} {045047} (\bibinfo {year}
  {2019})}\BibitemShut {NoStop}%
\bibitem [{\citenamefont {Han}\ \emph {et~al.}(2015)\citenamefont {Han},
  \citenamefont {V.},\ and\ \citenamefont {Shklovskii}}]{ref20}%
  \BibitemOpen
  \bibfield  {author} {\bibinfo {author} {\bibfnamefont {F.}~\bibnamefont
  {Han}}, \bibinfo {author} {\bibfnamefont {R.~K.}\ \bibnamefont {V.}},\ and\
  \bibinfo {author} {\bibfnamefont {B.~I.}\ \bibnamefont {Shklovskii}},\ }\href
  {https://doi.org/10.1103/PhysRevB.92.035204} {\bibfield  {journal} {\bibinfo
  {journal} {Phys. Rev. B}\ }\textbf {\bibinfo {volume} {92}},\ \bibinfo
  {pages} {035204} (\bibinfo {year} {2015})}\BibitemShut {NoStop}%
\bibitem [{\citenamefont {M.}\ and\ \citenamefont
  {G{\"u}{{\c{c}}}l{\"u}}(2020)}]{ref21}%
  \BibitemOpen
  \bibfield  {author} {\bibinfo {author} {\bibfnamefont {P.}~\bibnamefont
  {M.}}\ and\ \bibinfo {author} {\bibfnamefont {A.~D.}\ \bibnamefont
  {G{\"u}{{\c{c}}}l{\"u}}},\ }\href
  {https://doi.org/10.1103/PhysRevB.102.174204} {\bibfield  {journal} {\bibinfo
   {journal} {Phys. Rev. B}\ }\textbf {\bibinfo {volume} {102}},\ \bibinfo
  {pages} {174204} (\bibinfo {year} {2020})}\BibitemShut {NoStop}%
\bibitem [{\citenamefont {Polat}\ \emph {et~al.}(2020)\citenamefont {Polat},
  \citenamefont {Sevin{{\c{c}}}li},\ and\ \citenamefont
  {G{\"u}{\c{c}}l{\"u}}}]{ref22}%
  \BibitemOpen
  \bibfield  {author} {\bibinfo {author} {\bibfnamefont {M.}~\bibnamefont
  {Polat}}, \bibinfo {author} {\bibfnamefont {H.}~\bibnamefont
  {Sevin{{\c{c}}}li}},\ and\ \bibinfo {author} {\bibfnamefont {A.~D.}\
  \bibnamefont {G{\"u}{\c{c}}l{\"u}}},\ }\href
  {https://doi.org/10.1103/PhysRevB.101.205429} {\bibfield  {journal} {\bibinfo
   {journal} {Phys. Rev. B}\ }\textbf {\bibinfo {volume} {101}},\ \bibinfo
  {pages} {205429} (\bibinfo {year} {2020})}\BibitemShut {NoStop}%
\bibitem [{\citenamefont {Gamayun}\ \emph {et~al.}(2009)\citenamefont
  {Gamayun}, \citenamefont {Gorbar},\ and\ \citenamefont {Gusynin}}]{ref23}%
  \BibitemOpen
  \bibfield  {author} {\bibinfo {author} {\bibfnamefont {O.~V.}\ \bibnamefont
  {Gamayun}}, \bibinfo {author} {\bibfnamefont {E.~V.}\ \bibnamefont
  {Gorbar}},\ and\ \bibinfo {author} {\bibfnamefont {V.~P.}\ \bibnamefont
  {Gusynin}},\ }\href {https://doi.org/10.1103/PhysRevB.80.165429} {\bibfield
  {journal} {\bibinfo  {journal} {Phys. Rev. B}\ }\textbf {\bibinfo {volume}
  {80}},\ \bibinfo {pages} {165429} (\bibinfo {year} {2009})}\BibitemShut
  {NoStop}%
\bibitem [{\citenamefont {Pereira}\ \emph {et~al.}(2008)\citenamefont
  {Pereira}, \citenamefont {Kotov},\ and\ \citenamefont {Castro~Neto}}]{ref24}%
  \BibitemOpen
  \bibfield  {author} {\bibinfo {author} {\bibfnamefont {V.~M.}\ \bibnamefont
  {Pereira}}, \bibinfo {author} {\bibfnamefont {V.~N.}\ \bibnamefont {Kotov}},\
  and\ \bibinfo {author} {\bibfnamefont {A.~H.}\ \bibnamefont {Castro~Neto}},\
  }\href {https://doi.org/10.1103/PhysRevB.78.085101} {\bibfield  {journal}
  {\bibinfo  {journal} {Phys. Rev. B}\ }\textbf {\bibinfo {volume} {78}},\
  \bibinfo {pages} {085101} (\bibinfo {year} {2008})}\BibitemShut {NoStop}%
\bibitem [{\citenamefont {Zhu}\ \emph {et~al.}(2014)\citenamefont {Zhu},
  \citenamefont {Liu},\ and\ \citenamefont {Yang}}]{ref25}%
  \BibitemOpen
  \bibfield  {author} {\bibinfo {author} {\bibfnamefont {J.~L.}\ \bibnamefont
  {Zhu}}, \bibinfo {author} {\bibfnamefont {C.}~\bibnamefont {Liu}},\ and\
  \bibinfo {author} {\bibfnamefont {N.}~\bibnamefont {Yang}},\ }\href
  {https://doi.org/10.1103/PhysRevB.90.125405} {\bibfield  {journal} {\bibinfo
  {journal} {Phys. Rev. B}\ }\textbf {\bibinfo {volume} {90}},\ \bibinfo
  {pages} {125405} (\bibinfo {year} {2014})}\BibitemShut {NoStop}%
\bibitem [{\citenamefont {Downing}\ and\ \citenamefont
  {Portnoi}(2014)}]{ref26}%
  \BibitemOpen
  \bibfield  {author} {\bibinfo {author} {\bibfnamefont {C.~A.}\ \bibnamefont
  {Downing}}\ and\ \bibinfo {author} {\bibfnamefont {M.~E.}\ \bibnamefont
  {Portnoi}},\ }\href {https://doi.org/10.1103/PhysRevA.90.052116} {\bibfield
  {journal} {\bibinfo  {journal} {Phys. Rev. A}\ }\textbf {\bibinfo {volume}
  {90}},\ \bibinfo {pages} {052116} (\bibinfo {year} {2014})}\BibitemShut
  {NoStop}%
\bibitem [{\citenamefont {Pottelberge}\ \emph
  {et~al.}(2018{\natexlab{b}})\citenamefont {Pottelberge}, \citenamefont
  {Zarenia},\ and\ \citenamefont {Peeters}}]{ref27}%
  \BibitemOpen
  \bibfield  {author} {\bibinfo {author} {\bibfnamefont {R.~V.}\ \bibnamefont
  {Pottelberge}}, \bibinfo {author} {\bibfnamefont {M.}~\bibnamefont
  {Zarenia}},\ and\ \bibinfo {author} {\bibfnamefont {F.~M.}\ \bibnamefont
  {Peeters}},\ }\href {https://doi.org/10.1103/PhysRevB.97.207403} {\bibfield
  {journal} {\bibinfo  {journal} {Phys. Rev. B}\ }\textbf {\bibinfo {volume}
  {97}},\ \bibinfo {pages} {207403} (\bibinfo {year}
  {2018}{\natexlab{b}})}\BibitemShut {NoStop}%
\bibitem [{\citenamefont {Moldovan}\ \emph {et~al.}(2017)\citenamefont
  {Moldovan}, \citenamefont {Andelkovic},\ and\ \citenamefont
  {Peeters}}]{ref28}%
  \BibitemOpen
  \bibfield  {author} {\bibinfo {author} {\bibfnamefont {D.}~\bibnamefont
  {Moldovan}}, \bibinfo {author} {\bibfnamefont {M.}~\bibnamefont
  {Andelkovic}},\ and\ \bibinfo {author} {\bibfnamefont {F.~M.}\ \bibnamefont
  {Peeters}},\ }\href {https://doi.org/10.5281/zenodo.826942} {\bibinfo {title}
  {Pybinding v0.9.4: A python package for tight-binding calculations}}
  (\bibinfo {year} {2017})\BibitemShut {NoStop}%
\bibitem [{\citenamefont {Son}\ \emph {et~al.}(2006)\citenamefont {Son},
  \citenamefont {Cohen},\ and\ \citenamefont {Louie}}]{ref29}%
  \BibitemOpen
  \bibfield  {author} {\bibinfo {author} {\bibfnamefont {Y.~W.}\ \bibnamefont
  {Son}}, \bibinfo {author} {\bibfnamefont {M.~L.}\ \bibnamefont {Cohen}},\
  and\ \bibinfo {author} {\bibfnamefont {S.~G.}\ \bibnamefont {Louie}},\ }\href
  {https://doi.org/10.1103/PhysRevLett.97.216803} {\bibfield  {journal}
  {\bibinfo  {journal} {Phys. Rev. Lett.}\ }\textbf {\bibinfo {volume} {97}},\
  \bibinfo {pages} {216803} (\bibinfo {year} {2006})}\BibitemShut {NoStop}%
\bibitem [{\citenamefont {Han}\ \emph {et~al.}(2007)\citenamefont {Han},
  \citenamefont {{\"O}zyilmaz}, \citenamefont {Zhang},\ and\ \citenamefont
  {Kim}}]{ref30}%
  \BibitemOpen
  \bibfield  {author} {\bibinfo {author} {\bibfnamefont {M.~Y.}\ \bibnamefont
  {Han}}, \bibinfo {author} {\bibfnamefont {B.}~\bibnamefont {{\"O}zyilmaz}},
  \bibinfo {author} {\bibfnamefont {Y.}~\bibnamefont {Zhang}},\ and\ \bibinfo
  {author} {\bibfnamefont {P.}~\bibnamefont {Kim}},\ }\href
  {https://doi.org/10.1103/PhysRevLett.98.206805} {\bibfield  {journal}
  {\bibinfo  {journal} {Phys. Rev. Lett}\ }\textbf {\bibinfo {volume} {98}},\
  \bibinfo {pages} {206805} (\bibinfo {year} {2007})}\BibitemShut {NoStop}%
\bibitem [{\citenamefont {Orlof}\ \emph {et~al.}(2013)\citenamefont {Orlof},
  \citenamefont {Ruseckas},\ and\ \citenamefont {Zozoulenko}}]{ref31}%
  \BibitemOpen
  \bibfield  {author} {\bibinfo {author} {\bibfnamefont {A.}~\bibnamefont
  {Orlof}}, \bibinfo {author} {\bibfnamefont {J.}~\bibnamefont {Ruseckas}},\
  and\ \bibinfo {author} {\bibfnamefont {I.~V.}\ \bibnamefont {Zozoulenko}},\
  }\href {https://doi.org/10.1103/PhysRevB.88.125409} {\bibfield  {journal}
  {\bibinfo  {journal} {Phys. Rev. B}\ }\textbf {\bibinfo {volume} {88}},\
  \bibinfo {pages} {125409} (\bibinfo {year} {2013})}\BibitemShut {NoStop}%
\bibitem [{\citenamefont {S.}\ and\ \citenamefont {Schmelcher}(2012)}]{ref32}%
  \BibitemOpen
  \bibfield  {author} {\bibinfo {author} {\bibfnamefont {M.~B.}\ \bibnamefont
  {S.}}\ and\ \bibinfo {author} {\bibfnamefont {P.}~\bibnamefont
  {Schmelcher}},\ }\href {https://doi.org/10.1103/PhysRevB.86.245404}
  {\bibfield  {journal} {\bibinfo  {journal} {Phys. Rev. B}\ }\textbf {\bibinfo
  {volume} {86}},\ \bibinfo {pages} {245404} (\bibinfo {year}
  {2012})}\BibitemShut {NoStop}%
\bibitem [{\citenamefont {Pottelberge}\ \emph
  {et~al.}(2018{\natexlab{c}})\citenamefont {Pottelberge}, \citenamefont
  {Zarenia},\ and\ \citenamefont {Peeters}}]{ref33}%
  \BibitemOpen
  \bibfield  {author} {\bibinfo {author} {\bibfnamefont {R.~V.}\ \bibnamefont
  {Pottelberge}}, \bibinfo {author} {\bibfnamefont {M.}~\bibnamefont
  {Zarenia}},\ and\ \bibinfo {author} {\bibfnamefont {F.~M.}\ \bibnamefont
  {Peeters}},\ }\href {https://doi.org/10.1103/PhysRevB.98.115406} {\bibfield
  {journal} {\bibinfo  {journal} {Phys. Rev. B}\ }\textbf {\bibinfo {volume}
  {98}},\ \bibinfo {pages} {115406} (\bibinfo {year}
  {2018}{\natexlab{c}})}\BibitemShut {NoStop}%
\bibitem [{\citenamefont {Pottelberge}\ \emph {et~al.}(2017)\citenamefont
  {Pottelberge}, \citenamefont {Zarenia}, \citenamefont {Vasilopoulos},\ and\
  \citenamefont {Peeters}}]{ref34}%
  \BibitemOpen
  \bibfield  {author} {\bibinfo {author} {\bibfnamefont {R.~V.}\ \bibnamefont
  {Pottelberge}}, \bibinfo {author} {\bibfnamefont {M.}~\bibnamefont
  {Zarenia}}, \bibinfo {author} {\bibfnamefont {P.}~\bibnamefont
  {Vasilopoulos}},\ and\ \bibinfo {author} {\bibfnamefont {F.~M.}\ \bibnamefont
  {Peeters}},\ }\href {https://doi.org/10.1103/PhysRevB.95.245410} {\bibfield
  {journal} {\bibinfo  {journal} {Phys. Rev. B}\ }\textbf {\bibinfo {volume}
  {95}},\ \bibinfo {pages} {245410} (\bibinfo {year} {2017})}\BibitemShut
  {NoStop}%
\bibitem [{\citenamefont {Jiao}\ \emph {et~al.}(2009)\citenamefont {Jiao},
  \citenamefont {Zhang}, \citenamefont {Wang}, \citenamefont {Diankov},\ and\
  \citenamefont {Dai}}]{ref35}%
  \BibitemOpen
  \bibfield  {author} {\bibinfo {author} {\bibfnamefont {L.}~\bibnamefont
  {Jiao}}, \bibinfo {author} {\bibfnamefont {L.}~\bibnamefont {Zhang}},
  \bibinfo {author} {\bibfnamefont {X.}~\bibnamefont {Wang}}, \bibinfo {author}
  {\bibfnamefont {G.}~\bibnamefont {Diankov}},\ and\ \bibinfo {author}
  {\bibfnamefont {H.}~\bibnamefont {Dai}},\ }\href
  {https://doi.org/10.1038/nature07919} {\bibfield  {journal} {\bibinfo
  {journal} {Nature}\ }\textbf {\bibinfo {volume} {458}},\ \bibinfo {pages}
  {877} (\bibinfo {year} {2009})}\BibitemShut {NoStop}%
\bibitem [{\citenamefont {Tapasztó}\ \emph {et~al.}(2008)\citenamefont
  {Tapasztó}, \citenamefont {Dobrik}, \citenamefont {Lambin},\ and\
  \citenamefont {Biró}}]{ref36}%
  \BibitemOpen
  \bibfield  {author} {\bibinfo {author} {\bibfnamefont {L.}~\bibnamefont
  {Tapasztó}}, \bibinfo {author} {\bibfnamefont {G.}~\bibnamefont {Dobrik}},
  \bibinfo {author} {\bibfnamefont {P.}~\bibnamefont {Lambin}},\ and\ \bibinfo
  {author} {\bibfnamefont {L.~P.}\ \bibnamefont {Biró}},\ }\href
  {https://doi.org/10.1038/nnano.2008.149} {\bibfield  {journal} {\bibinfo
  {journal} {Nat. Nanotech.}\ }\textbf {\bibinfo {volume} {3}},\ \bibinfo
  {pages} {397} (\bibinfo {year} {2008})}\BibitemShut {NoStop}%
\bibitem [{\citenamefont {Sprinkle}\ \emph {et~al.}(2010)\citenamefont
  {Sprinkle}, \citenamefont {Ruan}, \citenamefont {Hu}, \citenamefont
  {Hankinson}, \citenamefont {Roy}, \citenamefont {Zhang}, \citenamefont {Wu},
  \citenamefont {Berger},\ and\ \citenamefont {Heer}}]{ref37}%
  \BibitemOpen
  \bibfield  {author} {\bibinfo {author} {\bibfnamefont {M.}~\bibnamefont
  {Sprinkle}}, \bibinfo {author} {\bibfnamefont {M.}~\bibnamefont {Ruan}},
  \bibinfo {author} {\bibfnamefont {Y.}~\bibnamefont {Hu}}, \bibinfo {author}
  {\bibfnamefont {J.}~\bibnamefont {Hankinson}}, \bibinfo {author}
  {\bibfnamefont {M.~R.}\ \bibnamefont {Roy}}, \bibinfo {author} {\bibfnamefont
  {B.}~\bibnamefont {Zhang}}, \bibinfo {author} {\bibfnamefont
  {X.}~\bibnamefont {Wu}}, \bibinfo {author} {\bibfnamefont {C.}~\bibnamefont
  {Berger}},\ and\ \bibinfo {author} {\bibfnamefont {W.~A.~d.}\ \bibnamefont
  {Heer}},\ }\href {https://doi.org/10.1038/nnano.2010.192} {\bibfield
  {journal} {\bibinfo  {journal} {Nat. Nanotech.}\ }\textbf {\bibinfo {volume}
  {5}},\ \bibinfo {pages} {727} (\bibinfo {year} {2010})}\BibitemShut {NoStop}%
\bibitem [{\citenamefont {Liu}\ \emph {et~al.}(2014)\citenamefont {Liu},
  \citenamefont {Kim}, \citenamefont {Hsu}, \citenamefont {Sokolov},
  \citenamefont {Yap}, \citenamefont {Yuan}, \citenamefont {W.}, \citenamefont
  {H.}, \citenamefont {Y.}, \citenamefont {Hwang},\ and\ \citenamefont
  {Bao}}]{ref38}%
  \BibitemOpen
  \bibfield  {author} {\bibinfo {author} {\bibfnamefont {N.}~\bibnamefont
  {Liu}}, \bibinfo {author} {\bibfnamefont {K.}~\bibnamefont {Kim}}, \bibinfo
  {author} {\bibfnamefont {P.}~\bibnamefont {Hsu}}, \bibinfo {author}
  {\bibfnamefont {A.~N.}\ \bibnamefont {Sokolov}}, \bibinfo {author}
  {\bibfnamefont {F.~L.}\ \bibnamefont {Yap}}, \bibinfo {author} {\bibfnamefont
  {H.~T.}\ \bibnamefont {Yuan}}, \bibinfo {author} {\bibfnamefont {X.~Y.}\
  \bibnamefont {W.}}, \bibinfo {author} {\bibfnamefont {Y.}~\bibnamefont {H.}},
  \bibinfo {author} {\bibfnamefont {C.}~\bibnamefont {Y.}}, \bibinfo {author}
  {\bibfnamefont {H.~Y.}\ \bibnamefont {Hwang}},\ and\ \bibinfo {author}
  {\bibfnamefont {Z.}~\bibnamefont {Bao}},\ }\href
  {https://doi.org/10.1021/ja509871n} {\bibfield  {journal} {\bibinfo
  {journal} {J. Am. Chem. Soc.}\ }\textbf {\bibinfo {volume} {136}},\ \bibinfo
  {pages} {17284} (\bibinfo {year} {2014})}\BibitemShut {NoStop}%
\bibitem [{\citenamefont {Linden}\ \emph {et~al.}(2012)\citenamefont {Linden},
  \citenamefont {Zhong}, \citenamefont {Timmer}, \citenamefont {Aghdassi},
  \citenamefont {Franke}, \citenamefont {Zhang}, \citenamefont {Feng},
  \citenamefont {Mllen}, \citenamefont {Fuchs}, \citenamefont {Chi},\ and\
  \citenamefont {Zacharias}}]{ref39}%
  \BibitemOpen
  \bibfield  {author} {\bibinfo {author} {\bibfnamefont {S.}~\bibnamefont
  {Linden}}, \bibinfo {author} {\bibfnamefont {D.}~\bibnamefont {Zhong}},
  \bibinfo {author} {\bibfnamefont {A.}~\bibnamefont {Timmer}}, \bibinfo
  {author} {\bibfnamefont {N.}~\bibnamefont {Aghdassi}}, \bibinfo {author}
  {\bibfnamefont {J.~H.}\ \bibnamefont {Franke}}, \bibinfo {author}
  {\bibfnamefont {H.}~\bibnamefont {Zhang}}, \bibinfo {author} {\bibfnamefont
  {X.}~\bibnamefont {Feng}}, \bibinfo {author} {\bibfnamefont {K.}~\bibnamefont
  {Mllen}}, \bibinfo {author} {\bibfnamefont {H.}~\bibnamefont {Fuchs}},
  \bibinfo {author} {\bibfnamefont {L.}~\bibnamefont {Chi}},\ and\ \bibinfo
  {author} {\bibfnamefont {H.}~\bibnamefont {Zacharias}},\ }\href
  {https://doi.org/10.1103/PhysRevLett.108.216801} {\bibfield  {journal}
  {\bibinfo  {journal} {Phys. Rev. Lett.}\ }\textbf {\bibinfo {volume} {108}},\
  \bibinfo {pages} {216801} (\bibinfo {year} {2012})}\BibitemShut {NoStop}%
\bibitem [{\citenamefont {Munkhbat}\ \emph {et~al.}(2020)\citenamefont
  {Munkhbat}, \citenamefont {Yankovich}, \citenamefont {Baranov}, \citenamefont
  {Verre}, \citenamefont {Olsson},\ and\ \citenamefont {Shegai}}]{ref40}%
  \BibitemOpen
  \bibfield  {author} {\bibinfo {author} {\bibfnamefont {B.}~\bibnamefont
  {Munkhbat}}, \bibinfo {author} {\bibfnamefont {A.~B.}\ \bibnamefont
  {Yankovich}}, \bibinfo {author} {\bibfnamefont {D.~G.}\ \bibnamefont
  {Baranov}}, \bibinfo {author} {\bibfnamefont {R.}~\bibnamefont {Verre}},
  \bibinfo {author} {\bibfnamefont {E.}~\bibnamefont {Olsson}},\ and\ \bibinfo
  {author} {\bibfnamefont {T.~O.}\ \bibnamefont {Shegai}},\ }\href
  {https://doi.org/10.1038/S41467-020-18428-2} {\bibfield  {journal} {\bibinfo
  {journal} {Nat. Commun.}\ }\textbf {\bibinfo {volume} {11}},\ \bibinfo
  {pages} {4604} (\bibinfo {year} {2020})}\BibitemShut {NoStop}%
\end{thebibliography}%


%

\end{document}